\documentclass[draftcls, onecolumn]{IEEEtran}

\usepackage{amssymb}
\usepackage{amsthm}
\usepackage{amsfonts}
\usepackage{verbatim}
\usepackage{pstricks}

\usepackage[textwidth=2cm,textsize=footnotesize]{todonotes}
\usepackage[permil]{overpic}

\usepackage[ruled,vlined]{algorithm2e}
\usepackage[T1]{fontenc}
\usepackage{cite}
\usepackage{amsmath}
\usepackage{psfrag}
\usepackage{amsbsy}
\usepackage{graphicx}
\usepackage{setspace}
\usepackage[nomarkers,nolists]{endfloat}

\def\MMSE{{\text{MMSE}}}
\DeclareMathOperator*{\argmax}{arg\,max}

\begin{document}

\title{Framework for Link-Level Energy Efficiency Optimization with Informed Transmitter}

\author{Christian~Isheden,~\IEEEmembership{Member,~IEEE,}
	Zhijiat~Chong,~\IEEEmembership{Member,~IEEE,}
	Eduard~Jorswieck,~\IEEEmembership{Senior Member,~IEEE,}
	and~Gerhard~Fettweis,~\IEEEmembership{Fellow,~IEEE}
\thanks{C. Isheden was with the Vodafone Chair Mobile Communications Systems, Technische Universit\"at Dresden, D-01069 Dresden, Germany, and is currently with Actix GmbH, D-01067 Dresden, Germany, e-mail: christian.isheden@actix.com. }
\thanks{Z. Chong and E. Jorswieck are with the Chair of Communications Theory, Technische Universit\"at Dresden.}
\thanks{G. Fettweis is with the Vodafone Chair Mobile Communications Systems, Technische Universit\"at Dresden.}
}

\maketitle

\begin{abstract}

The dramatic increase of network infrastructure comes at the cost of rapidly increasing energy consumption, which makes optimization of energy efficiency (EE) an important topic.
Since EE is often modeled as the ratio of rate to power, we present a mathematical framework called fractional programming that provides insight into this class of optimization problems, as well as algorithms for computing the solution. The main idea is that the objective function is transformed to a weighted sum of rate and power. A generic problem formulation for systems dissipating transmit-independent circuit power in addition to transmit-dependent power is presented. We show that a broad class of EE maximization problems can be solved efficiently, provided the rate is a concave function of the transmit power. We elaborate examples of various system models including time-varying parallel channels. Rate functions with an arbitrary discrete modulation scheme are also treated. The examples considered lead to water-filling solutions, but these are different from the dual problems of power minimization under rate constraints and rate maximization under power constraints, respectively, because the constraints need not be active. We also demonstrate that if the solution to a rate maximization problem is known, it can be utilized to reduce the EE problem into a one-dimensional convex problem. 

\end{abstract}
\section{Introduction}
\label{sec:Introduction}

Exponentially increasing data traffic and demand for ubiquitous access have triggered a dramatic expansion of network infrastructure, which comes at the cost of rapidly increasing energy consumption and a considerable carbon footprint of the mobile communications industry. Therefore, increasing the energy efficiency (EE) in cellular networks has become an important and urgent task.
Apart from this, EE plays an important role in other areas of wireless communications as well. For example, in multihop networks, EE is critical for prolonging the lifetime of the network \cite{bae2009end}.
EE is also becoming increasingly important  in mobile communication devices since battery capacity
is unable to keep pace with increasing power dissipation of signal processing circuits \cite{Pentikousis2010}.

A comprehensive survey of joint PHY and MAC layer techniques for improving wireless EE can be found in \cite{miao2009cross}. In an effort to integrate the fundamental issues related to EE in wireless networks, \cite{Chen2011b} presents four fundamental EE trade-offs in detail. The paper at hand is concerned with the trade-off between spectral efficiency (SE) and EE. In particular, we look at practical transmission systems dissipating transmit-independent circuit power in addition to the transmit-dependent power. As described in \cite{Prabhu2010}, the link-level EE optimization problem in the active mode is closely related to two classical problems, one being rate maximization subject to a maximum power constraint and the other one being power minimization subject to a minimum rate constraint. Both problems lead to water-filling solutions, with the water level determined by the respective constraint. In comparison, EE optimization involves maximizing the amount of transmitted data per unit energy, or equivalently minimizing the energy consumption per bit. It turns out that the EE optimization problem also results in water-filling solutions, with a water level that depends on the transmit-independent power.
 Results on energy-efficient link adaptation for frequency-selective channels are presented in \cite{miao2010energy}. 
 A related efficiency objective function, which involves the packet success rate, has been treated in a game-theoretic setting utilizing pricing to achieve EE in \cite{Meshkati2007}. 

The contribution of this paper is a framework for solving EE maximization problems, which are different from the related problems of power minimization under rate constraints and rate maximization under power constraints, respectively. 
EE maximization belongs to a class of optimization problems called fractional programs. Since the fractional programming theory is not well-known in the wireless communications community, results that are presently scattered in the operations research literature are summarized in a coherent manner. With this, we also show that the various approaches to the problem are mathematically connected through a scalarized bi-criterion optimization problem and provide an efficient solution algorithm. These results can be used to solve a large class of EE problems based on various system models.
A series of applications ranging from time-invariant, flat-fading parallel channels to time-varying, flat-fading (single and parallel) channels illustrates the applicability of the developed framework. Results are shown to be applicable even for discrete modulation schemes. The algorithmic solutions have very low complexity because they are based on water-filling power allocation. In contrast to sum rate maximization or sum power minimization, however, the water level is not adjusted iteratively to satisfy the constraint with equality. Instead, the water level is used as a parameter that is adjusted until a certain criterion corresponding to the maximum EE is fulfilled. Finally, a direct reuse of standard rate maximization algorithms in a nested programming procedure, which is made possible using the framework, is discussed. 
 
 

The outline is as follows. A motivating example including the channel and power model is given in Section \ref{sec:motivating_example}. Section \ref{sec:fracprog} lays out the mathematical framework for the paper. Both the maximization case (for maximizing the bit/J metric), and the minimization case (for minimizing the J/bit) are discussed. Incorporation of various empirical power dissipation models into a generic EE problem formulation is demonstrated in Section \ref{sec:powermodels}. Based on this generic problem, results for different fading models (static and time-varying channels) and for practical modulation schemes are presented in Section \ref{sec:channel_models}. We further discuss how known rate maximization algorithms from the literature can be adopted to EE optimization. Simulation results based on the models discussed are presented in Section \ref{sec:simulation}. The paper is wrapped up with a discussion about the water-filling solutions in Section \ref{sec:discussion}, followed by some conclusions in Section \ref{sec:conclusions}.


Our notation is as follows. Column vectors are denoted by bold lowercase letters, \emph{e.g.} $\boldsymbol{x}$, with the \emph{i}th component denoted by $x_i$. 
 Sets are denoted by calligraphic letters such as $\mathcal{S}$. 
 $\left[x\right]^{+}$ denotes $\max\left\{ 0,x\right\}$. 
 $\left[x\right]^{y}_{z}$ denotes $\min \left\{ y, \max\left\{ z,x\right\} \right\}$.
 A column vector of all ones is denoted by $\boldsymbol{1}$, and the component-wise sum of a vector $\boldsymbol{x}$ is denoted by $\boldsymbol{1}^T \boldsymbol{x}$.

\section{Motivating example}
\label{sec:motivating_example}

In order to motivate the development of a general framework, we provide an anecdotal example of EE maximization. Consider a time-invariant Gaussian channel with $K$ parallel quasi-static block flat-fading channels with coherence time $T_c$ and gains $\gamma_1,...,\gamma_K$. Perfect channel state information (CSI) is available at the transmitter as well as at the receiver. Each parallel channel occupies a bandwidth of $W_c$ and elastic data is to be transmitted. Assuming Gaussian codebooks at the transmit side, the achievable data rate on channel $k$ in bits per complex dimension is $r_k(p_k) = \log_2 (1 + \gamma_k p_k)$ with transmit power allocation per unit bandwidth $p_k \geq 0$. The amount of information transmitted during a time-frequency chunk $T_c W_c$ is given by 
\begin{eqnarray}
	T_c W_c \sum_{k=1}^K \log_2 (1 + \gamma_k p_k) \quad \text{[bits]} \label{eq:f1}
\end{eqnarray}

In \cite{Cui2004}, a power model for the nodes in a wireless network is proposed. The total power consumption in the active mode at the transmitter is modeled as $P_\textup{ont} = P_\textup{PA} + P_\textup{ct},$ where  $P_\textup{PA}$ is the power dissipated in the power amplifier and $P_\textup{ct}$ is the power dissipated in all other circuit blocks. The power dissipated in the power amplifier is given by $P_\textup{PA} = \frac{\xi}{\eta} P_t,$ where $\xi$ and $\eta$ are the power amplifier output backoff (OBO) and drain efficiency, respectively, and $P_t=W_c \sum_{k=1}^K p_k$ is the transmit power. The OBO is needed to avoid the nonlinear region of the power amplifier and is determined by the peak-to-average power ratio (PAPR). The circuit power $P_\textup{ct}$ is given by $P_\textup{ct} = P_\textup{mix} + P_\textup{syn} + P_\textup{filt} + P_\textup{DAC},$ where the terms correspond to the power dissipation of the mixer, the frequency synthesizer,  the active filters, and the digital-to-analog converter, respectively. The amount of energy consumed during one time-frequency chunk is 
\begin{eqnarray}
 T_c \cdot \left( P_{ct} + P_{PA} \right) = T_c W_c \frac{\xi}{\eta} \left( \mu + \sum_{k=1}^K p_k \right) \quad \text{[Joule]}\label{eq:f2},
\end{eqnarray}
where $\mu = \frac{\eta}{\xi} \frac{P_\textup{ct}}{W_c}$ [W/Hz]. 

In a general sense, efficiency can be seen as the extent to which a resource, such as electricity, is used for the intended purpose. Efficiency is a measurable concept, quantitatively determined by the ratio of output to input. In the physical and medium access control layers, the output is the effective amount of data transmitted (measured in bits or nats) and the input is the total energy consumed for transmitting the data (in Joule). This results in the \emph{EE}, defined as the amount of data transmitted (\ref{eq:f1}) divided by the amount of energy consumed (\ref{eq:f2}) as  
\begin{eqnarray}
	EE = \log_2 e \cdot \frac{\eta}{\xi} \frac{ \sum_{k=1}^K \log (1 + \gamma_k p_k)}{ \mu + \sum_{k=1}^K p_k} = \log_2 e \cdot \frac{\eta}{\xi} \frac{f_1(\boldsymbol{p})}{ f_2(\boldsymbol{p})} \quad \text{[bits/Joule]} \label{eq:EEexam}. 
\end{eqnarray}
The EE in (\ref{eq:EEexam}) is usually maximized subject to constraints on the transmit powers $p_1,...,p_K$ and the sum rate. Spectral mask constraints $0 \leq p_k \leq p_{max}$ are required by regulatory bodies. Sum power constraints are required in order to limit interference in neighboring sectors. An additional sum rate constraint $R_0$ can model the quality of service requirement of the traffic in the next block. Based on (\ref{eq:EEexam}), the resulting optimization problem is  
\begin{eqnarray}
	\underset{\boldsymbol{p} \in \mathcal{S}}{\text{maximize}} \; \frac{f_1(\boldsymbol{p})}{ f_2(\boldsymbol{p})} \quad \text{with} \quad \mathcal{S} = \{ \boldsymbol{p} \in \mathbb{R}_+^K : \sum p_k \leq P, p_k \leq p_{max}, f_1(\boldsymbol{p}) \geq R_0 \} \label{eq:noo}, 
\end{eqnarray}
where $P$ is the maximum sum power. Problem (\ref{eq:noo}) belongs to a class of optimization problems called fractional programs. As we will see later, many more examples of EE maximization problems in different wireless communication scenarios lead to fractional programs. Therefore, we study this class in more detail in the next section.

\section{Fractional Programming}
\label{sec:fracprog}

Fractional programs are nonlinear programs where the objective function is a ratio of two real-valued functions. For simplicity, only differentiable fractional programs, \emph{i.e.} where both the numerator and the denominator are differentiable, are considered in this section. A general nonlinear fractional program has the form
\begin{equation}
  \begin{array}{ll}
  \underset{\boldsymbol{x} \in \mathcal{S}}{\text{maximize}} & q(\boldsymbol{x}) = \frac{f_1(\boldsymbol{x})}{f_2(\boldsymbol{x})}, \\
  \end{array}
  \label{eq:ConcaveConvex}
\end{equation}
where $\mathcal{S} \subseteq \mathbb{R}^n$, $f_1, f_2: \mathcal{S} \rightarrow \mathbb{R}$ and $f_2(\boldsymbol{x}) > 0$. Problem \eqref{eq:ConcaveConvex} is called a concave-convex fractional program if $f_1$ is concave, $f_2$ is convex, and $\mathcal{S}$ is a convex set; additionally $f_1(\boldsymbol{x}) \geq 0$ is required, unless $f_2$ is affine. When $f_1$ and $f_2$ are differentiable, the objective function in \eqref{eq:ConcaveConvex} is pseudoconcave \cite{Schaible1983}, implying that any stationary point is a global maximum and that the Karush-Kuhn-Tucker (KKT) conditions are sufficient if a constraint qualification is fulfilled. Because of this, \eqref{eq:ConcaveConvex} can be solved directly by various convex programming algorithms \cite{Schaible1983}. 
However, when $f_1$ is concave and $f_2$ is convex, the fractional program can be transformed to an equivalent convex program, which may be solved more efficiently in certain cases. In the literature, two different convex formulations and an approach based on duality have been suggested \cite{Schaible1983b}. In the following, we will discuss each approach in some detail. As we will see, however, they are very closely related since they all lead to the same optimality condition.

\subsection{Parametric convex program}
\label{sec:par-convex-prog}

Consider the following equivalent form \cite[p. 134]{Boyd2004} of the fractional program \eqref{eq:ConcaveConvex}:
$$
  \begin{array}{ll}
  \underset{\boldsymbol{x} \in \mathcal{S}, \lambda \in \mathbb{R}}{\text{maximize}} & \lambda \\
  \text{subject to} &  \frac{f_1(\boldsymbol{x})}{f_2(\boldsymbol{x})} - \lambda \geq 0
  \end{array}
$$
Rearranging the constraint, we obtain
$$
  \begin{array}{ll}
  \underset{\boldsymbol{x} \in \mathcal{S}, \lambda \in \mathbb{R}}{\text{maximize}} & \lambda \\
  \text{subject to} &  f_1(\boldsymbol{x}) - \lambda f_2(\boldsymbol{x}) \geq 0.
  \end{array}
$$
This formulation is not jointly convex in $\boldsymbol{x}$ and $\lambda$, but for a fixed value of $\lambda$ we have a feasibility problem in $\boldsymbol{x}$, which is convex if $f_1$ is concave and $f_2$ is convex. The problem is feasible if \[ \underset{\boldsymbol{x} \in \mathcal{S}}{\text{max}} \quad f_1(\boldsymbol{x}) - \lambda f_2(\boldsymbol{x}) \geq 0.\] One can use bisection to find the optimal value of the parameter $\lambda$, solving the feasibility problem at each step of the algorithm, as described in more detail in \cite[pp. 145-146]{Boyd2004}.

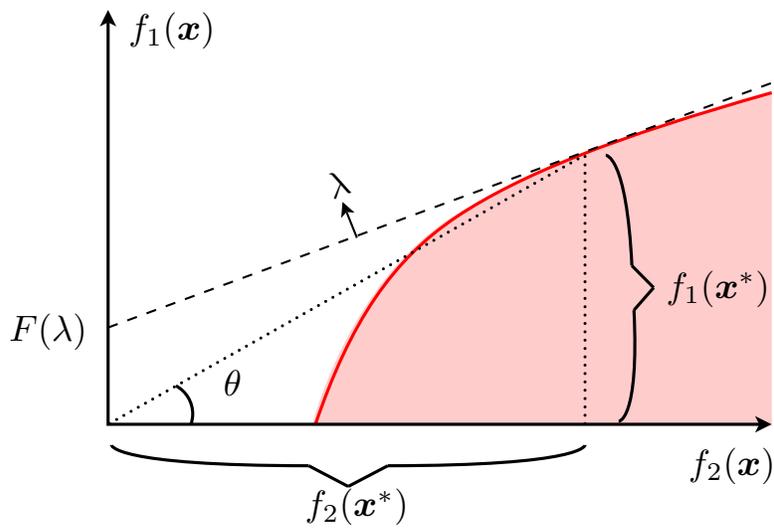
\begin{figure}
    \centering
    \resizebox{0.9\columnwidth}{!}{
\psset{unit=1mm}
\psset{linewidth=0.3,dotsep=1,hatchwidth=0.3,hatchsep=1.5,shadowsize=1,dimen=middle}
\psset{dotsize=0.7 2.5,dotscale=1 1,fillcolor=black}
\psset{arrowsize=1 2,arrowlength=1,arrowinset=0.25,tbarsize=0.7 5,bracketlength=0.15,rbracketlength=0.15}
\begin{pspicture}(0,0)(106.58,54.84)
\newrgbcolor{userLineColour}{1 0.8 0.8}
\newrgbcolor{userFillColour}{1 0.8 0.8}
\pscustom[linecolor=userLineColour,fillcolor=userFillColour,fillstyle=solid]{\psbezier(27.87,10)(27.87,10)(31.26,23.58)(41.47,30.02)
\psbezier(51.67,36.46)(71.87,42)(71.87,42)
\psline(71.87,42)(71.87,10)
\psbezier(71.87,10)(71.87,10)(71.87,10)
\psline(71.87,10)(27.87,10)
\closepath}
\rput(-2,6){}
\rput[l](10,48){$f_1(\boldsymbol{x})$}
\pscustom[linecolor=red]{\psbezier(28,10)(31.36,19.89)(35.44,25.9)(41.6,30.02)
\psbezier(47.76,34.13)(56.88,37.73)(72,42)
}
\psline{<->}(72,10)
(8,10)(8,50)
\psline[linestyle=dotted,dotsep=0.5](8,10)(54,36)
\rput(7.71,53.63){}
\rput[l](64,6){$f_2(\boldsymbol{x})$}
\rput{0}(13.09,11){\psellipticarc[](0,0)(3.08,3.08){-18.95}{61.64}}
\rput(106.58,54.84){}
\psline[linewidth=0.2,linestyle=dashed,dash=1 1](8,19.34)(71.92,42.92)
\rput{22}(30.13,33.21){$\lambda$}
\rput[r](6,18.95){$F(\lambda)$}
\psline[linestyle=dotted,dotsep=0.5](54,36)(54,10)
\psline[linewidth=0.2]{->}(32,28)(30.61,31.47)
\rput[l](61.87,23.08){$f_1(\boldsymbol{x}^*)$}
\rput(32,2){$f_2(\boldsymbol{x}^*)$}
\psbezier(34.98,6)(54,6)(54,8)(54,8)
\psbezier(27.35,6)(8.33,6)(8.33,8)(8.33,8)
\psline(27.35,6)(31.16,4)
\psline(31.16,4)(34.98,6)
\psbezier(58.52,25.31)(58,36)(56,35.97)(56,35.97)
\psbezier(58.72,21.02)(59.24,10.32)(57.24,10.29)(57.24,10.29)
\psline(58.72,21.02)(60.62,23.19)
\psline(60.62,23.19)(58.52,25.31)
\rput(20,14){$\theta$}
\end{pspicture}
}
    \caption{Illustration of the trade-off curve between $f_1(\boldsymbol{x}^*)$ and $f_2(\boldsymbol{x}^*)$, where $\boldsymbol{x}^*$ is optimal for a given value of $\lambda$. The parameter $\lambda$ is the slope of the tangent, whereas $F(\lambda)$ is given by the intersection with the vertical axis. The corresponding value of the objective function in \eqref{eq:ConcaveConvex} is given by $\tan \theta$. The maximum occurs where $F(\lambda) = 0$.}
    \label{fig:pareto}
\end{figure}

Consider the function
\begin{equation}
  F(\lambda) = \underset{\boldsymbol{x} \in \mathcal{S}}{\text{max}} \quad f_1(\boldsymbol{x}) - \lambda f_2(\boldsymbol{x}).
  \label{eq:F_lambda}
\end{equation}
It can be shown that $F(\lambda)$ is convex, continuous and strictly decreasing in $\lambda$ \cite{Dinkelbach1967}. The right hand side of \eqref{eq:F_lambda} can be viewed as a scalarized bi-criterion optimization problem in which $f_1(\boldsymbol{x})$ is to be maximized whereas $f_2(\boldsymbol{x})$ is to be minimized. The parameter $\lambda$ determines the relative weight of the denominator. If $\boldsymbol{x}^*$ is optimal for the scalar problem, then it is Pareto-optimal for the bi-criterion optimization problem \cite[pp. 178-184]{Boyd2004}. The set of Pareto optimal values for a bi-criterion problem is called the optimal trade-off curve. By varying the value of $\lambda$, we explore the optimal trade-off curve between the objectives, as illustrated in Figure \ref{fig:pareto}. The slope of the optimal trade-off curve at any point represents the local optimal trade-off between the two objectives. Where the slope is steep, small changes in $f_2$ result in large changes in $f_1$. The intersection of the curve with a vertical line $f_2 = \alpha$ gives the maximum value of $f_1$ that achieves $f_2 \leq \alpha$. Similarly, the intersection with a horizontal line $f_1 = \beta$ gives the minimum value of $f_2$ that achieves $f_1 \geq \beta$. 

Let $q^*$ be the optimum value of the objective function in \eqref{eq:ConcaveConvex}. The following statements are equivalent\footnote{In fact, these properties of $F(\lambda)$ are true for more general nonlinear fractional programs \cite{Dinkelbach1967}.} \cite{Schaible1983b}:
\begin{align*}
F(\lambda) > 0 & \Leftrightarrow \lambda < q^* \\
F(\lambda) = 0 & \Leftrightarrow \lambda = q^* \\
F(\lambda) < 0 & \Leftrightarrow \lambda > q^*
\end{align*}
Thus, solving problem \eqref{eq:ConcaveConvex} is equivalent to finding the
root of the nonlinear function $F(\lambda)$, so the condition for optimality is
\begin{equation}
  F(\lambda^*) = \underset{\boldsymbol{x} \in \mathcal{S}}{\text{max}} \quad f_1(\boldsymbol{x}) - \lambda^* f_2(\boldsymbol{x}) = 0.
  \label{eq:Cond1}
\end{equation}


Various iterative algorithms are available for finding the root of $F(\lambda)$ \cite{ibaraki1983parametric}. For example, the Dinkelbach method \cite{Dinkelbach1967} in Algorithm \ref{alg:Dinkelbach} is based on the application of Newton's method. To see this, note that the update in Newton's method is calculated as \[\lambda_{n+1} = \lambda_n - \frac{F(\lambda_n)}{F'(\lambda_n)} = \lambda_n - \frac{f_1(\boldsymbol{x}_n^*) - \lambda_n f_2(\boldsymbol{x}_n^*)}{-f_2(\boldsymbol{x}_n^*)} = \frac{f_1(\boldsymbol{x}_n^*)}{f_2(\boldsymbol{x}_n^*)}.\] Therefore,
the sequence converges to the optimal point with a superlinear
convergence rate. A detailed convergence analysis can be found in
\cite{Schaible1976a}. The initial point can be any $\lambda_0$
that satisfies $F(\lambda_0) \geq 0$.
\begin{figure}
    \begin{algorithm}[H]
    \SetAlgoLined
        \KwData{$\lambda_0$ satisfying $F(\lambda_0) \geq 0$, tolerance $\Delta$}
        $n = 0$ \;
        \While{$|F(\lambda_n)| \geq \Delta$}{ 
        Use $\lambda = \lambda_n$ in \eqref{eq:F_lambda} to obtain $\boldsymbol{x}_n^*$\;
        $\lambda_{n+1} = \frac{f_1(\boldsymbol{x}_n^*)}{f_2(\boldsymbol{x}_n^*)}$\;
        $n++$\;
        }
    \caption{The Dinkelbach method.}
    \label{alg:Dinkelbach}
    \end{algorithm}
\end{figure}
It is also straightforward to include box constraints for $f_1(\boldsymbol{x})$ or $f_2(\boldsymbol{x})$. Referring to Figure \ref{fig:pareto}, a lower bound on $f_1$ or $f_2$ corresponds to an upper bound on $\lambda$, say $\lambda_\text{max}$, whereas an upper bound on $f_1$ or $f_2$ corresponds to a lower bound $\lambda_\text{min}$. Therefore, solving an optimization problem with this kind of inequality constraints
reduces to solving the unconstrained problem and determining whether $\lambda^*$
falls within the interval $[\lambda_\text{min}, \lambda_\text{max}]$. If not,
$\lambda^*$ is replaced by the respective endpoint. 

\subsection{Parameter-free convex program}
\label{sec:ParFreeCvxProg}

Let $\mathcal{S}_0 \in \mathbb{R}^n$ be a nonempty, convex, and open subset of the domain of the objective function $q(\boldsymbol{x})$ that satisfies $f_2(\boldsymbol{x}) > 0$. Let $\mathcal{S} = \{\boldsymbol{x} \in \mathcal{S}_0 | \boldsymbol{g}(\boldsymbol{x}) \leq \boldsymbol{0}\}$ be the feasible subset of $\mathcal{S}_0$ with all convex inequality constraints $\boldsymbol{g}(\boldsymbol{x}) \leq \boldsymbol{0}$ taken into account.


The transformation 
\begin{equation}
  \boldsymbol{y} = \frac{1}{f_2(\boldsymbol{x})} \boldsymbol{x}, \quad t = \frac{1}{f_2(\boldsymbol{x})}, \quad \boldsymbol{x} \in \mathcal{S}_0
  \label{eq:transformation}
\end{equation}
yields the equivalent parameter-free problem \cite{Schaible1974}
\begin{equation}
  \begin{array}{ll}
  \underset{\boldsymbol{y}/t \in \mathcal{S}_0}{\text{maximize}} & t f_1\left(\boldsymbol{y}/t\right) \\
  \text{subject to} & t f_2\left(\boldsymbol{y}/t\right) \leq 1 \\
    & t \boldsymbol{g}\left(\boldsymbol{y}/t\right) \leq \boldsymbol{0},
  \end{array}
  \label{eq:Parameter-freeConvex}
\end{equation}
which is convex in $(\boldsymbol{y},t)$ since taking the perspective of a function preserves convexity. The inequality in the first constraint can be changed to an equality if $f_2(\boldsymbol{x})$ is affine. Problem \eqref{eq:ConcaveConvex} has an optimal solution if and only if problem \eqref{eq:Parameter-freeConvex} has one, and the solutions are related by \eqref{eq:transformation}.

Let the dual variables associated with the constraints $t f_2\left(\boldsymbol{y}/t\right) - 1 \leq 0$ and $t \boldsymbol{g}\left(\boldsymbol{y}/t\right) \leq \boldsymbol{0}$ be denoted by $\lambda$ and $\boldsymbol{u}$, respectively.
The Lagrangian is \[\mathcal{L}(\boldsymbol{y},t,\lambda,\boldsymbol{u}) = -t f_1\left(\boldsymbol{y}/t\right) + \lambda \left(t f_2\left(\boldsymbol{y}/t\right) - 1\right) + \left(t \boldsymbol{g}\left(\boldsymbol{y}/t\right)\right)^T \boldsymbol{u}\] and the resulting stationarity conditions are
\[\left\{ 
  \begin{array}{ll}
  -\nabla f_1(\boldsymbol{y}^*/t^*) + \lambda^* \nabla f_2(\boldsymbol{y}^*/t^*) + (\nabla \boldsymbol{g}(\boldsymbol{y}^*/t^*))^T \boldsymbol{u}^* &= \boldsymbol{0} \\
  -f_1(\boldsymbol{y}^*/t^*) + \lambda^*  f_2(\boldsymbol{y}^*/t^*) + \left(\boldsymbol{g}\left(\boldsymbol{y}^*/t^*\right)\right)^T \boldsymbol{u}^* &= 0.
  \end{array}
\right. \] 
Due to complementary slackness, the last term in the second row is zero. The first row is the condition for the maximum of $f_1(\boldsymbol{y}/t) - \lambda^* f_2(\boldsymbol{y}/t)$ subject to $\boldsymbol{y}/t \in \mathcal{S}$ with $\lambda^*$ as parameter. Thus, the condition for the optimum is 
\begin{equation}
	F(\lambda^*) = \underset{\boldsymbol{y}/t \in \mathcal{S}}{\text{max}} \quad f_1(\boldsymbol{y}/t) - \lambda^* f_2(\boldsymbol{y}/t) = 0.
	\label{eq:parfreeF}
\end{equation}
Comparing this to \eqref{eq:Cond1}, we see that the resulting optimality condition is equivalent to the one in the parametric approach.

\subsection{Dual program}

The Wolfe dual of the equivalent convex program \eqref{eq:Parameter-freeConvex} is (after substituting $\boldsymbol{x}$ for $\boldsymbol{y}/t$) \cite{schaible1976fractional}
\begin{equation}
  \begin{array}{ll}
  \text{minimize} & \lambda \\
  \text{subject to} & -\nabla f_1(\boldsymbol{x}) + \lambda \nabla f_2(\boldsymbol{x}) + (\nabla \boldsymbol{g}(\boldsymbol{x}))^T \boldsymbol{u} = \boldsymbol{0} \\
	& -f_1(\boldsymbol{x}) + \lambda f_2(\boldsymbol{x}) + (\boldsymbol{g}(\boldsymbol{x}))^T \boldsymbol{u} \geq 0 \\
	& \boldsymbol{x} \in \mathcal{S}_0, \boldsymbol{u} \in \mathbb{R}_+^m, \lambda \geq 0,
  \end{array}
  \label{eq:dual}
\end{equation}
which coincides \cite{Schaible1974} with the dual of the parametric convex program
\begin{equation}
  \begin{array}{ll}
  \underset{\boldsymbol{x} \in \mathcal{S}}{\text{maximize}} & f_1(\boldsymbol{x}) - \lambda f_2(\boldsymbol{x}), \\
  \end{array}
  \label{eq:ParamProb}
\end{equation}
where $\lambda \in \mathbb{R}$ is treated as a parameter. Thus, \eqref{eq:dual} is the dual of both convex programs. Note that the dual problem is not convex in general, since the equality constraint is typically not affine.


Based on Wolfe's direct duality theorem we have the following result \cite{Schaible1974}:
If $\boldsymbol{x}^*$ is an optimal solution to problem \eqref{eq:ConcaveConvex} and $\mathcal{S}$ is nonempty, then there are $\boldsymbol{u}^*$ and $\lambda^*$ such that $(\boldsymbol{x}^*, \boldsymbol{u}^*, \lambda^*)$ is an optimal solution to the dual problem \eqref{eq:dual} and $q(\boldsymbol{x}^*) = \lambda^*$.

At the optimum, the inequality in the dual problem is satisfied with equality, \emph{i.e.} $-f_1(\boldsymbol{x}^*) + \lambda^* f_2(\boldsymbol{x}^*) + (\boldsymbol{g}(\boldsymbol{x}^*))^T \boldsymbol{u}^* = 0$. Since $\lambda^* = f_1(\boldsymbol{x}^*)/f_2(\boldsymbol{x}^*)$, due to complementary slackness we have $(\boldsymbol{g}(\boldsymbol{x}^*))^T \boldsymbol{u}^* = 0$. Thus, problem \eqref{eq:dual} reduces to finding $\boldsymbol{x}^*$ and the optimal Lagrange multiplier $\lambda^*$ such that 
\[\left\{ 
  \begin{array}{ll}
        -\nabla f_1(\boldsymbol{x}^*) + \lambda^* \nabla f_2(\boldsymbol{x}^*) + (\nabla \boldsymbol{g}(\boldsymbol{x}^*))^T \boldsymbol{u} &= \boldsymbol{0} \\
        -f_1(\boldsymbol{x}^*) + \lambda^* f_2(\boldsymbol{x}^*) &= 0 .
  \end{array} 
\right. \] 
The first equation is the condition for the maximum of $f_1(\boldsymbol{x}) - \lambda^* f_2(\boldsymbol{x})$ over $\boldsymbol{x} \in \mathcal{S}$, with $\lambda^*$ as parameter. Summarizing, the condition for the optimum is \[F(\lambda^*) = \underset{\boldsymbol{x} \in \mathcal{S}}{\text{max}} \quad f_1(\boldsymbol{x}) - \lambda^* f_2(\boldsymbol{x}) = 0.\]
Again, this is equivalent to \eqref{eq:Cond1}.

\subsection{Convex fractional program}

Here we consider the equivalent convex-concave minimization problem with convex inequality constraints. In this case, we have
$$
  \begin{array}{ll}
  \underset{\boldsymbol{x} \in \mathcal{S}}{\text{minimize}} & \frac{f_2(\boldsymbol{x})}{f_1(\boldsymbol{x})}, \\
  \end{array}
$$
where $\mathcal{S} =\{\boldsymbol{x} \in \mathcal{S}_0 | \boldsymbol{g}(\boldsymbol{x}) \leq \boldsymbol{0}\}$ is bounded, and where $g_i(\boldsymbol{x})$ is convex and differentiable, $f_1(\boldsymbol{x})$ is nonnegative, concave, and $f_2(\boldsymbol{x})$ is positive, convex on $\mathcal{S}$.

Consider the epigraph form of the convex fractional program:
$$
  \begin{array}{ll}
  \underset{\boldsymbol{x} \in \mathcal{S}, \tilde{\lambda} \in \mathbb{R}}{\text{minimize}} & \tilde{\lambda} \\
  \text{subject to} &  \frac{f_2(\boldsymbol{x})}{f_1(\boldsymbol{x})} - \tilde{\lambda} \leq 0
  \end{array}
$$
Rearranging the constraint, we obtain
$$
  \begin{array}{ll}
  \underset{\boldsymbol{x} \in \mathcal{S}, \tilde{\lambda} \in \mathbb{R}}{\text{minimize}} & \tilde{\lambda} \\
  \text{subject to} &  f_2(\boldsymbol{x}) - \tilde{\lambda} f_1(\boldsymbol{x}) \leq 0.
  \end{array}
$$
This formulation is not jointly convex, but for a given value of $\tilde{\lambda}$ we have a convex feasibility problem in $\boldsymbol{x}$. The feasibility problem is solved by minimizing $f_2(\boldsymbol{x}) - \tilde{\lambda} f_1(\boldsymbol{x})$ and determining if the result is less than or equal to zero. Note further that the constraint must be active at the optimum, so we have \[\underset{\boldsymbol{x} \in \mathcal{S}}{\text{min}} \quad f_2(\boldsymbol{x}) - \tilde{\lambda}^* f_1(\boldsymbol{x}) = 0.\]

The dual problem is given by \cite{jagannathan1973duality}
$$
  \begin{array}{ll}
  \text{maximize} & \tilde{\lambda} \\
  \text{subject to} & \nabla f_2(\boldsymbol{x}) - \tilde{\lambda} \nabla f_1(\boldsymbol{x}) + (\nabla \boldsymbol{g}(\boldsymbol{x}))^T \boldsymbol{u} = 0 \\
	& f_2(\boldsymbol{x}) - \tilde{\lambda} f_1(\boldsymbol{x})+ (\boldsymbol{g}(\boldsymbol{x}))^T \boldsymbol{u} \geq 0 \\
	& \boldsymbol{x} \in \mathcal{S}, \boldsymbol{u} \in \mathbb{R}_+^m, \tilde{\lambda} \geq 0,
  \end{array}
$$
which is analogous to \eqref{eq:dual}.

\section{Power models for base stations}
\label{sec:powermodels}

As described in Section \ref{sec:motivating_example}, we are interested in maximizing the ratio of achievable rate to dissipated power, where the power consists of a transmit-independent part in addition to the total transmit power. We will concentrate on the generic optimization problem
\begin{equation}
  \begin{array}{ll}
  \underset{\boldsymbol{p} \in \mathcal{S}}{\text{maximize}} & q(\boldsymbol{p}) = \frac{r(\boldsymbol{p})}{\mu + \sum_{i=1}^K p_i}, \\
  \end{array}
  \label{eq:BasicProb}
\end{equation}
where $\boldsymbol{p}$ is the transmit power spectral density, $r(\boldsymbol{p})$ is a general concave rate (spectral efficiency in nat/s) function, and $\mu > 0$ is a constant offset, corresponding to the relative weight of the transmit-independent power. The optimal value of the objective function decreases when $\mu$ increases, because $\mu$ corresponds to a shift to the right of the curve in Fig. \ref{fig:pareto}. In this section, it will be demonstrated that various EE maximization problems resulting from power models in the literature can be transformed to the generic problem form \eqref{eq:BasicProb}. While these power models are all linear, the framework in this paper allows for arbitrary convex functions of transmit power.


\subsection{Generic base station power model}
\label{sec:genericBS}

In \cite{Chen2011}, a generic model for the total power consumption of a base station $P_\text{tot}$ is suggested, based on the assumptions
\begin{enumerate}
\item the total transmit power $P_{t}$ is equally allocated to the $n_a$ antennas at the transmitter,
\item each antenna is associated with an RF chain, including a power amplifier, $P_{PA}$, and other RF hardware, $P_c$,
\item the power dissipation of each PA is considered proportional to the output power, $P_{PA} = P_t/n_a/\eta_{PA}$.
\end{enumerate}
The model is
\[P_\text{tot} = \frac{\frac{P_t}{\eta_{PA}} + n_a P_c + P_\text{sta}}{\eta_{PS}\left(1-\eta_C\right)},\] where $P_\text{sta}$ is the static power consumption from baseband processing and battery unit, $\eta_{PS}$ is the efficiency of the power supply, and $\eta_C$ is the efficiency loss in the cooling system. 


The EE metric [in bit/J] can be written as \[EE = \frac{B \cdot r(p)}{ C \cdot \eta_{PA} \cdot B (\mu + p)} = \frac{q(p)}{C \cdot \eta_{PA}},\] where $B$ is the system bandwidth, $p = P_{t}/B,$ $\mu = \eta_{PA} \left( n_a P_c + P_\text{sta}\right)/B,$ and $C = \eta_{PS}\left(1-\eta_C\right).$

\subsection{Macro base station power model}

In \cite{Arnold2010}, the following power model for macro and micro base stations is presented:
\[
    P_\textup{BS} = N_\textup{Sector} \cdot N_\textup{PApSec}
    \cdot \left( \frac{P_\textup{TX}}{\mu_\textup{PA}} + P_\textup{SP}
    \right) \cdot \left( 1+C_\textup{C} \right)
    \cdot \left( 1+C_\textup{PSBB} \right)
\]
The main parameters in the model for a macro basestation are summarized in Table~\ref{tab:table_parameter}.

\begin{table}
\renewcommand{\arraystretch}{1.3}
\caption{Linear power model parameters} \label{tab:table_parameter} \centering
\begin{tabular}{c|l}
\hline
\small \bfseries Parameter & \small \bfseries Description \\
\hline\hline
$N_\textup{Sector}$ & \# sectors \\
$P_\textup{TX}$ & Tx power \\
$P_\textup{SP}$ & Signal processing overhead \\
$C_\textup{PSBB}$ & Battery backup and power supply loss \\
$N_\textup{PApSec}$ & \# PAs per sector\\
$\mu_\textup{PA}$ & PA efficiency\\
$C_\textup{C}$ & Cooling loss\\
\hline
\end{tabular}
\vspace{-15pt}
\end{table}

With $p = P_\textup{TX} / B,$ $\mu = P_\textup{SP} \cdot \mu_\textup{PA} / B,$ and $C = N_\textup{Sector} \cdot N_\textup{PApSec} \cdot  \left( 1+C_\textup{C} \right) \cdot \left( 1+C_\textup{PSBB} \right),$ 
the EE metric is \[EE = \frac{B \cdot r(p)}{ C \cdot \mu_\textup{PA} \cdot B (\mu + p)} = \frac{1}{C \cdot \mu_\textup{PA}} \cdot q(p).\]

\section{Applications}
\label{sec:channel_models}

In this section, we shall demonstrate how various channel models (flat fading and frequency-selective fading, static and time-variant), antenna configurations (including SISO and MIMO), and input constellations (Gaussian and quadratic M-QAM) result in concave rate functions that can all be treated within the mathematical framework developed thus far.

	\subsection{Time-invariant parallel subchannels}
\label{sec:TimeInvariantParChan}

From Section \ref{sec:motivating_example}, the problem to be solved is
\begin{equation}
  \begin{array}{ll}
  \underset{\boldsymbol{p} \in \mathcal{S}}{\text{maximize}} & q(\boldsymbol{p}) = \frac{\boldsymbol{1}^T \boldsymbol{r}(\boldsymbol{p})}{\mu + \boldsymbol{1}^T \boldsymbol{p}}, \\
  \end{array}
  \label{eq:OptimProblemOrig_fs}
\end{equation}
where $r_i(p_i) = \log\left(1 + \gamma_i p_i\right)$. Here, $\gamma_i = \frac{|h_i|^2}{N_0}$ is the channel-to-noise ratio (CNR) of subchannel $i$. Furthermore, we have box constraints for the individual powers, $0 \leq p_i \leq p_\text{max}$, $i=1, \ldots, K.$ Thus, the feasible set $\mathcal{S}$ is compact (closed and bounded) and convex. In order to illustrate the fractional programming theory, we shall solve problem \eqref{eq:OptimProblemOrig_fs} using both the parametric and the parameter-free approach.

\subsubsection{Parametric convex problem}

The function $F(\lambda)$ is given by
\begin{equation}
  F(\lambda) = \underset{\boldsymbol{p} \in \mathcal{S}}{\text{max}} \quad \boldsymbol{1}^T \boldsymbol{r}(\boldsymbol{p}) - \lambda (\mu + \boldsymbol{1}^T \boldsymbol{p}). \\
  \label{eq:ConvProb_fs}
\end{equation}
%
The stationarity condition is \[\frac{d r_i}{d p_i}\bigg|_{p_i=p_i^*} - \lambda = 0, \quad i = 1, \ldots, K.\] Thus, we have 
\[\lambda = \frac{\gamma_i}{1 + \gamma_i p_i^*}.\]
Taking the box constraints into account, the optimal power allocation is
\begin{equation}
  p_i^*(\lambda) = \left[\frac{1}{\lambda} -
\frac{1}{\gamma_i}\right]_0^{p_\text{max}}.
  \label{eq:P_fs}
\end{equation}
The parameter $\lambda$ corresponds to a cutoff CNR. A subcarrier is not
used if its CNR falls below the cutoff value
($\gamma_i < \lambda$). The optimal power is therefore given by
water-filling. 

The explicit solution in \eqref{eq:P_fs} is used in every iteration of any method that finds the root of $F(\lambda)$. 
One way of finding the root is to use the Dinkelbach method, as shown in Algorithm \ref{alg:Dinkelbach_fs}.

\begin{figure}%
	    \begin{algorithm}[H]
    	\SetAlgoLined
        \KwData{$\lambda_0$ satisfying $F(\lambda_0) \geq 0$, tolerance $\Delta$}
        $n = 0$\;
        \While{$|F(\lambda_n)| \geq \Delta$}{ 
        Calculate $\boldsymbol{p}_n^*$ from \eqref{eq:P_fs}\;
        $\lambda_{n+1} = \frac{\boldsymbol{1}^T \boldsymbol{r}(\boldsymbol{p}_n^*)}{\mu + \boldsymbol{1}^T \boldsymbol{p}_n^*}$\;
        $n++$\;
        }
    \caption{The Dinkelbach method for energy-efficient link adaptation on a block fading, frequency-selective channel as modeled by optimization problem \eqref{eq:OptimProblemOrig_fs}.}
    \label{alg:Dinkelbach_fs}
\end{algorithm}
\end{figure}

Referring to Fig. \ref{fig:pareto}, the vertical axis corresponds to the sum rate, whereas the horizontal axis corresponds to sum power plus an offset $\mu$. A point on the trade-off curve corresponds to water-filling with a given water level $1/\lambda$. A point below the curve corresponds to a sub-optimal power distribution. The curve crosses the horizontal axis at $\mu$ and the optimal EE occurs where the tangent goes through the origin. When the offset increases, the optimal EE decreases, and it occurs for a higher sum power.

\subsubsection{Parameter-free convex problem}

Remember that $\mathcal{S} = \{\boldsymbol{x} \in \mathcal{S}_0 | \boldsymbol{g}(\boldsymbol{x}) \leq \boldsymbol{0}\}$, where $\mathcal{S}_0$ is the part of the domain of the objective function whose denominator is positive. In our case, the domain can be characterized as follows: The logarithmic function is only defined for the positive real domain, which implies $p_i > -1/\gamma_i$, and the denominator cannot be zero, so $\boldsymbol{1}^T \boldsymbol{p} \neq -\mu$. The requirement that the denominator be positive excludes all vectors with a sum less than or equal to $-\mu$.

By the transformation \[\boldsymbol{y} = \frac{1}{\mu + \boldsymbol{1}^T \boldsymbol{p}} \boldsymbol{p}; \quad t = \frac{1}{\mu + \boldsymbol{1}^T \boldsymbol{p}}; \quad \boldsymbol{p} \in \mathcal{S}_0,\] we obtain the convex problem
$$
  \begin{array}{lll}
  \underset{\boldsymbol{y}/t \in \mathcal{S}_0}{\text{maximize}} & t \boldsymbol{1}^T \boldsymbol{r}(\boldsymbol{y}/t) \\
  \text{subject to} & t \mu + \boldsymbol{1}^T \boldsymbol{y} = 1 \\
     & t \boldsymbol{g}\left(\boldsymbol{y}/t\right) \leq \boldsymbol{0},
  \end{array}
$$
where $r_i (y_i/t) = \log \left(1 + \gamma_i \cdot y_i/t\right)$ and $\boldsymbol{g}$ is a vector of box constraints $0 \leq y_i/t \leq p_{max}$. Here, the variable $t$ corresponds to the inverse of the total power dissipation.

After introduction of a Lagrange multiplier $\lambda \in \mathbb{R}$
for the equality constraint, the Lagrangian is
\[\mathcal{L}(\boldsymbol{y}, t, \lambda, \boldsymbol{u}) = -t \boldsymbol{1}^T \boldsymbol{r}(\boldsymbol{y}/t) + \lambda (t \mu + \boldsymbol{1}^T \boldsymbol{y} - 1) + \left(t \boldsymbol{g}\left(\boldsymbol{y}/t\right)\right)^T \boldsymbol{u}.\] As the reader can verify, the KKT conditions yield
\[ \frac{y_i^*}{t^*} = \left[\frac{1}{\lambda^*} - \frac{1}{\gamma_i}\right]_0^{p_\text{max}}, \quad i=1, \ldots, K,\] and $\lambda^* = t^* \boldsymbol{1}^T \boldsymbol{r}(\boldsymbol{y}^*/t^*).$

\subsubsection{Adding constraints}

\begin{figure}
    \centering
    \resizebox{0.9\columnwidth}{!}{\includegraphics{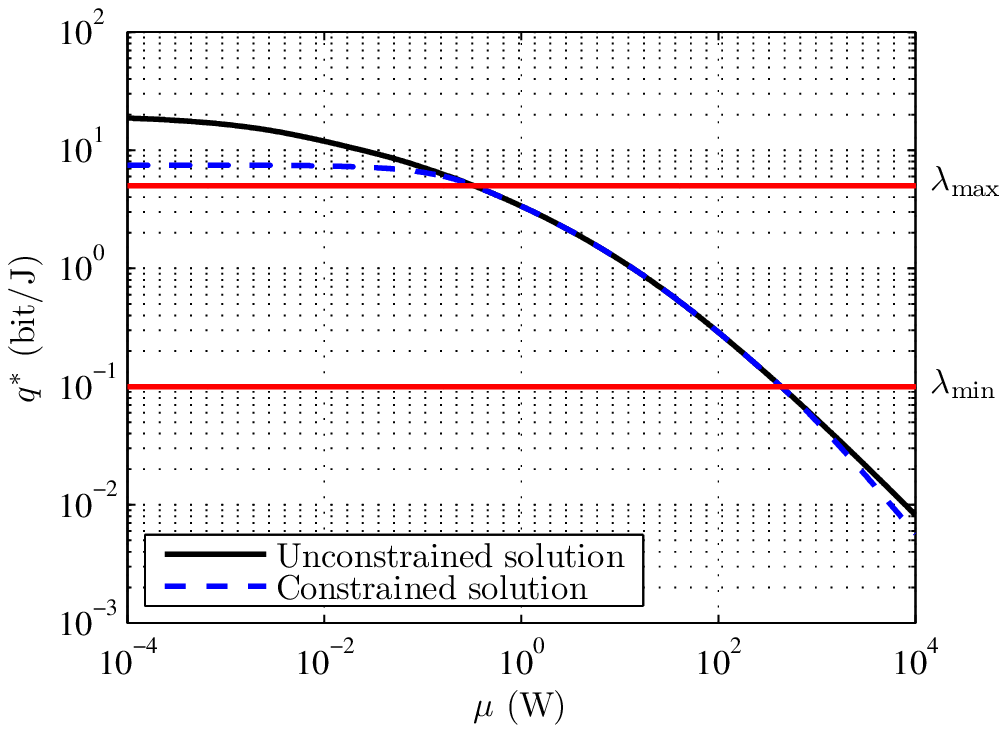}}
    \caption{Plot of the optimal EE as a function of $\mu$ for a frequency-selective channel. When $\mu$ is small, $\lambda = \lambda_\text{max}$ due to the sum rate constraint and the problem reduces to pure power minimization. Similarly, when $\mu$ is large, $\lambda = \lambda_\text{min}$ due to the maximum sum power constraint and the problem reduces to pure rate maximization. In both cases, there is a penalty in EE outside the interval.}
    \label{fig:constraints}
\end{figure}

As discussed previously, a maximum power constraint $\boldsymbol{1}^T \boldsymbol{p} \leq P$ corresponds to a lower bound $\lambda_\text{min}$ for $\lambda$. Similarly, a minimum rate constraint $\boldsymbol{1}^T \boldsymbol{r} \geq R$ corresponds to an upper bound $\lambda_\text{max}$. As illustrated in Figure \ref{fig:constraints}, these additional constraints lead to a penalty in EE.

\subsubsection{Flat fading channel}

For the flat fading channel, the optimal
power allocation reduces to
\begin{equation}
  p^*(\lambda) = \left[\frac{1}{\lambda} -
\frac{1}{\gamma}\right]_0^{p_\text{max}}.
  \label{eq:P_ff}
\end{equation}
For this simple channel model, it is in fact possible to derive the
optimal value $\lambda^*$ in closed form. Assume first that $\gamma \geq
\lambda^*$, so that $p^* \geq 0$. Again, we wish to find the
solution to the nonlinear equation $F(\lambda^*) = 0$, \emph{i.e.}
%
$$
  \log \frac{1}{\lambda^*} - \log \frac{1}{\gamma} - \lambda^* \left(\mu + \left(\frac{1}{\lambda^*} - \frac{1}{\gamma}\right)\right) = 0.
$$
%
After introduction of $s = \frac{\gamma}{\lambda^*}$, this can be transformed to \[(\log s - 1) \cdot s = \mu \gamma -
1.\] 
The solution to this equation is \[\log s = 1 +
W(e^{-1} (\mu \gamma - 1)),\] where $W$ is the Lambert W function
\cite{Corless1996}. Note that the condition $p^* \geq 0$ corresponds to
$s \geq 1$, which implies $W(e^{-1} (\mu \gamma - 1)) \geq -1$,
\emph{i.e.} the principal branch $W_0$ is selected.
Thus, \[\lambda^* = \frac{\gamma}{s} = \frac{\gamma}{\exp(1 + W_0(e^{-1}
(\mu \gamma - 1)))}.\] When there are no constraints on rate and
power, there is always a feasible solution.

Although the solution can be derived analytically for the flat-fading channel, it
may still be attractive to use the Dinkelbach method for
numerical evaluation, since evaluation of the Lambert W function also
relies on a root-finding algorithm.
	\subsection{Time-varying, flat-fading channel}

Here, we wish to maximize the average number of bits transmitted per unit energy consumed,
calculated as the ergodic capacity divided by the average dissipated
power. We assume causal CSI at the transmitter in an ideal case with zero-delay feedback which requires no additional power. The EE maximization problem can be stated as
\begin{equation}  
  \underset{p(\gamma) \geq 0}{\text{maximize}} \quad q[p(\gamma)] = \frac{\int_0^\infty \log (1 + \gamma p(\gamma)) f(\gamma) d\gamma}{\mu + \int_0^\infty p(\gamma) f(\gamma) d\gamma},  
  \label{eq:Optimproblem}
\end{equation}
where $f(\gamma)$ is the probability density function (PDF) of the fading distribution. Note that optimization problem \eqref{eq:Optimproblem} is concerned with finding an optimal function rather than a finite-dimensional vector as assumed in Section \ref{sec:fracprog}. However, the extension to optimization over functions is straightforward. The parametric convex optimization problem is
\begin{equation}  
  \underset{p(\gamma) \geq 0}{\text{maximize}} \int_0^\infty \log (1 + \gamma p(\gamma)) f(\gamma) d\gamma - \lambda \left(\mu + \int_0^\infty p(\gamma) f(\gamma) d\gamma\right),  
  \label{eq:Concaveprob}
\end{equation}
where $\lambda > 0$ is treated as a parameter.

Problem \eqref{eq:Concaveprob} needs to be solved in each step of
the Dinkelbach method. 
The stationarity condition (obtained by setting the functional
derivative with respect to $p$ equal to zero) is \[\frac{\gamma}{1 +
\gamma p^*(\gamma)} - \lambda = 0.\] Solving this equation for $p^*$, we get \[p^*
= \frac{1}{\lambda} - \frac{1}{\gamma}.\] The transmit power must be
nonnegative, so the solution is
\[ p^*(\lambda) = \left[\frac{1}{\lambda} -
\frac{1}{\gamma}\right]^+\]
and $\lambda$ corresponds to a cutoff CNR. Thus, we have
\begin{equation}
  F(\lambda) = \int_{\lambda}^\infty \log \left(\frac{\gamma}{\lambda}\right) f(\gamma) d\gamma - \lambda \left(\mu + \int_{\lambda}^\infty \left(\frac{1}{\lambda} - \frac{1}{\gamma}\right) f(\gamma) d\gamma\right).
  \label{eq:Fq}
\end{equation}
The solution to $F(\lambda) = 0$ must be found numerically because no
closed-form solutions exist for typical continuous distributions.
However, evaluating $F(\lambda)$ numerically for any given $\lambda$ is
straightforward. Therefore, the optimal value $\lambda^*$ can be found
iteratively.

If the instantaneous CNR is below the cutoff level, the optimal strategy at the transmitter is to be idle. The idle probability is calculated as
\begin{equation}
  \mathbb{P}(\gamma < \lambda^*) = 1 - \int_{\lambda^*}^\infty f(\gamma) d\gamma.
  \label{eq:Idleprob}
\end{equation}

\subsubsection{Adding constraints} \label{sec:TimeVaryFlatConstr}

A maximum power constraint $\bar{p} \leq
\bar{p}_\text{max}$ is equivalent to $\lambda \geq \lambda_\text{min}$, where $\lambda_\text{min}$
satisfies
\[\int_{\lambda_\text{min}}^\infty \left(\frac{1}{\lambda_\text{min}} -
\frac{1}{\gamma}\right) f(\gamma) d\gamma = \bar{p}_\text{max}.\]
Similarly, a minimum rate constraint $\bar{r}^* \geq \bar{r}_\text{min}$ is
equivalent to $\lambda \leq \lambda_\text{max}$, where $\lambda_\text{max}$ satisfies
\[\int_{\lambda_\text{max}}^\infty \log \left(\frac{\gamma}{\lambda_\text{max}}\right)
f(\gamma) d\gamma = \bar{r}_\text{min}.\] 


\subsubsection{Example: Rayleigh fading}

In Rayleigh fading, the PDF $f(\gamma)$ is
\cite{lee1990estimate}
\begin{equation}
  f(\gamma) = \frac{e^{-\gamma/\bar{\gamma}}}{\bar{\gamma}}.
  \label{eq:pdf_Rayleigh}
\end{equation}
The average CNR $\bar{\gamma}$ is given by $\bar{\gamma} = G \cdot
\frac{\sigma^2}{N_0},$ where $G$ is the path gain from the
transmitter to the receiver and $\sigma$ is the mean of the Rayleigh
distributed variable. Substituting \eqref{eq:pdf_Rayleigh} into
\eqref{eq:Fq} yields
\begin{equation*}
  F(\lambda) = \int_{\lambda}^\infty \log \left(\frac{\gamma}{\lambda}\right) \frac{e^{-\gamma/\bar{\gamma}}}{\bar{\gamma}} d\gamma - \lambda \left(\mu + \int_{\lambda}^\infty \left(\frac{1}{\lambda} - \frac{1}{\gamma}\right) \frac{e^{-\gamma/\bar{\gamma}}}{\bar{\gamma}} d\gamma\right).
\end{equation*}
After the variable transformation $x = \frac{\lambda}{\bar{\gamma}}$;
$t = \frac{\gamma}{\lambda}$ we obtain
\begin{equation*}
  F(\lambda) = \int_{1}^\infty \log t \cdot xe^{-xt} dt - \lambda \left(\mu + \frac{1}{\bar{\gamma}} \int_1^\infty \left(1 - t^{-1}\right) e^{-xt} dt\right),
\end{equation*}
where $x$ and $t$ are functions of $\lambda$. Through integration by parts, we have
\begin{equation*}
  \int_{1}^\infty \log t \cdot xe^{-xt} dt = \left[\log t \cdot \left(-e^{-xt}\right)\right]_1^\infty - \int_1^\infty \frac{1}{t} \left(-e^{-xt}\right) dt = \int_1^\infty \frac{1}{t} \cdot e^{-xt} dt = E_1(x),
\end{equation*}
where the generalized exponential integral $E_n(x)$ is defined by
\[E_n(x) = \int_1^\infty t^{-n} e^{-xt} dt, \quad x \geq 0.\]
Thus, \[F(\lambda) = E_1\left(\frac{\lambda}{\bar{\gamma}}\right) - \lambda
\left(\mu + \frac{1}{\bar{\gamma}}
\left(E_0\left(\frac{\lambda}{\bar{\gamma}}\right) -
E_1\left(\frac{\lambda}{\bar{\gamma}}\right)\right)\right),\] where
$E_0(x) = \int_1^\infty e^{-xt} dt = \frac{e^{-x}}{x}.$

The idle probability for Rayleigh fading is given by \[\mathbb{P}(\gamma <
\lambda^*) = 1 - \int_{\lambda^*}^\infty
\frac{e^{-\gamma/\bar{\gamma}}}{\bar{\gamma}} d\gamma = 1 - \frac{\lambda^*}{\bar{\gamma}} \cdot E_0\left(\frac{\lambda^*}{\bar{\gamma}}\right) = 1 - \exp\left(-\frac{\lambda^*}{\bar{\gamma}}\right).\]

	\subsection{Time-varying, parallel subchannels}
\label{sec:TVparchan}

 Suppose we have $K$ parallel channels, as in the case of frequency-selective multicarrier systems. 
 Additionally, the channels vary with time and the power allocation can be selected independently for every channel realization $\boldsymbol{\gamma}=(\gamma_1,...,\gamma_K)$. We can characterize the power allocation as the vector function of the channel realization $\boldsymbol{p}(\boldsymbol{\gamma})$.
 As previously discussed, we want to maximize the EE $q$, which is quantified here as the ratio of the ergodic capacity to the average dissipated power,  over vector function $\boldsymbol{p}(\boldsymbol{\gamma})$. 
 The maximization problem is then given as
\begin{equation}
	\underset{\boldsymbol{p}\left(\boldsymbol{\gamma}\right) \geq \boldsymbol{0}}{\mathrm{maximize}} 
		\quad q\left[\boldsymbol{p}\left(\boldsymbol{\gamma}\right)\right]=
\frac{
\displaystyle{\intop_{\boldsymbol{\gamma}\in \mathbb{R}_+^K}}\sum_{i=1}^{K}
		\log\left(1+\gamma_{i}p_{i}\left(\boldsymbol{\gamma}\right)\right)f(\boldsymbol{\gamma})
	d\boldsymbol{\gamma}}
	{\mu+
	\intop_{\boldsymbol{\gamma}\in \mathbb{R}_+^K} \sum_{i=1}^{K}
		p_{i}\left(\boldsymbol{\gamma}\right)f(\boldsymbol{\gamma})
	d\boldsymbol{\gamma}}, 
	\label{eq:timeVaryParProb}
\end{equation}
where $f(\boldsymbol{\gamma})=f(\gamma_1,...,\gamma_K)$ is the joint PDF of the $K$ subchannel CNRs. The corresponding parametric concave optimization problem with parameter $\lambda$ is 
\begin{equation}	
	\underset{\boldsymbol{p}\left(\boldsymbol{\gamma}\right) \geq \boldsymbol{0}}{\mathrm{maximize}} 
\int_{\boldsymbol{\gamma}\in \mathbb{R}_+^K}\sum_{i=1}^{K}
		\log\left(1+\gamma_{i}p_{i}\left(\boldsymbol{\gamma}\right)\right)f(\boldsymbol{\gamma})
	d\boldsymbol{\gamma}
	- \lambda \left(
	\mu+\int_{\boldsymbol{\gamma}\in \mathbb{R}_+^K}\sum_{i=1}^{K}
		p_{i}\left(\boldsymbol{\gamma}\right)f(\boldsymbol{\gamma})
	d\boldsymbol{\gamma} \right),
	\label{eq:TimeVaryParChanEE} 	
\end{equation}
which has to be solved at each step of the Dinkelbach method.
 Since the maximand of \eqref{eq:TimeVaryParChanEE} is a concave functional of $\boldsymbol{p}\left(\boldsymbol{\gamma}\right)$, the KKT conditions are sufficient for optimality.
 These conditions yield the optimal function
\begin{equation}
	p_{i}^{*}\left(\lambda,\boldsymbol{\gamma}\right) = 							
		\left[
			\frac{1}{\lambda}-\frac{1}{ \gamma_{i}}
					\right]^{+}, \quad i=1, \ldots, K.
	\label{eq:optimalfunctional}
\end{equation}
 Note that $p_i^*$ is an explicit expression of the component
$\gamma_{i}$ only and not of the vector $\boldsymbol{\gamma}$. 

 We now obtain the solution to \eqref{eq:timeVaryParProb} by finding $\lambda^{*}$ using \eqref{eq:optimalfunctional}.
 This is done by computing the root of the function
\begin{equation}
	F\left(\lambda\right) = 	
		\int_{\boldsymbol{\gamma}\in \mathbb{R}_+^K}
			\sum_{i=1}^K \log \left(
			1 + \gamma_{i}p_{i}^{*}\left(\lambda,\gamma_{i}\right)
															\right) 
			f(\boldsymbol{\gamma}) d\boldsymbol{\gamma}
																-\lambda\left(
	\mu + \int_{\boldsymbol{\gamma}\in \mathbb{R}_+^K}
		\sum_{i=1}^K p_{i}^{*}\left(\lambda,\gamma_{i}\right)
			f(\boldsymbol{\gamma}) d\boldsymbol{\gamma}
										\right)
\label{eq:rootproblem} 	
\end{equation}
using the Dinkelbach method.
 The computation of the integrals may be demanding, especially for $K>3$.
 However, the computation time can be reduced by exploiting the structure of $f(\boldsymbol{\gamma})$, \emph{e.g.} if the parallel subchannels are independent, $f(\boldsymbol{\gamma})$ can be written as a product of the PDFs of its components $\gamma_i$. 

 Analogously to Section \ref{sec:TimeVaryFlatConstr}, average sum power and sum rate constraints can be easily imposed here as well. Moreover, this method can be applied to MIMO channels, which are decomposed into parallel channels using singular-value decomposition \cite{Tse2006}. The case of Rayleigh fading channels has been treated in \cite{Chong2011c}.

	\subsection{Gap to capacity}

The Shannon capacity models the theoretically achievable rate for an ideal Gaussian input. In a real system, the achievable rate is often modeled using a gap depending on the modulation and coding schemes being used. In addition, a gap can be used to model the uncertainty in the received SNR.

\subsubsection{Constant gap to capacity}

The simplest variation of the rate function is to introduce a constant gap to capacity, as suggested in \cite{miao2010energy}. The rate function then becomes \[r_i(p_i) = \log\left(1 + \frac{\gamma_i}{\Gamma} \cdot p_i\right),\] where $\Gamma$ is the gap to capacity. Note that $\Gamma$ is independent of the subcarrier CNR. The simplest way of including such a gap is to exchange $\gamma_i$ for $\frac{\gamma_i}{\Gamma}$ in the water-filling solution.

\subsubsection{Subchannel-dependent gaps (mercury/water-filling)}
\label{sec:merc-waterf}

For an arbitrary modulation scheme, the rate function is described by the mutual information expression. In the following, the approach is described for parallel channels following \cite{Lozano2006}. It can be generalized to multiple antenna systems \cite{Perez-Cruz2010}.  

The input signals on the $i$-th channel $s_i$ (normalized with unit power) are from some modulation set $\mathcal{M}_i$, which can be discrete as well as continuous. The rate function is defined as the mutual information between input and output of the channel,
\begin{equation}
  r_i(p_i) = I(s_i; \sqrt{\gamma_i p_i} s_i + n_i), \quad \text{[nats/channel use]}
  \label{eq:mut_inf}
\end{equation}
 where $\gamma_i = |h_i|^2/\sigma^2$, $n_i$ is a zero-mean unit-variance proper complex Gaussian random variable and $p_i$ is the power allocated to the $i$-th channel.
 The mutual information in \eqref{eq:mut_inf} is strictly concave in $\boldsymbol{p}$ \cite[Appendix A]{Lozano2006}. In general, it is difficult to obtain a closed form expression for the mutual information. However, all optimization problems in the last section can be generalized by the following observation \cite{Guo2005}: If the signal-to-noise ratio on the $i$th channel is denoted by $\rho_i = \gamma_i p_i$, then 
\begin{equation}
  \frac{d}{d \rho_i} r_i(\rho_i) = \MMSE_i(\rho_i),
  \label{eq:derivative_MMSE}
\end{equation}
where $\MMSE_i(\rho_i) = \mathbb{E}_{s_i} [ |s_i - \hat{s}_i|^2 ]$ with MMSE estimate $\hat{s}_i = \mathbb{E}_{s_i}[ s_i | \sqrt{\rho_i} s_i + n_i = y_i]$. The MMSE is known in closed form for many important discrete and continuous constellations \cite[Section IV]{Lozano2006} and these expressions can be inserted into the KKT optimality conditions. In order to solve for the optimal power allocation, the inverse MMSE function $\MMSE_i^{-1}(\rho_i)$ is used. 

The parametric convex program is
$$
  \begin{array}{ll}
  F(\lambda) = \underset{\boldsymbol{p} \in \boldsymbol{S}}{\text{max}} & \boldsymbol{1}^T \boldsymbol{r}(\boldsymbol{p}) - \lambda (\mu + \boldsymbol{1}^T \boldsymbol{p}), \\
  \end{array}
$$
where $r_i = r_i(p_i)$ according to \eqref{eq:mut_inf} and $\lambda \in \mathbb{R}$ is treated as parameter. 
 The stationarity condition is 
\[
	\frac{d r_i}{d p_i}\bigg|_{p_i=p_i^*} - \lambda = 0, \quad i = 1, \ldots, K. 
	\]
 Inserting \eqref{eq:derivative_MMSE}, we have 
\[
	\gamma_i \MMSE_i(\gamma_i p_i^*) = \lambda , \quad i = 1, \ldots, K,
	\] 
\emph{i.e.} the MMSE of subchannel $i$ at the optimum power $p_i^*$ is given by \[\MMSE_i(\gamma_i p_i^*) = \frac{\lambda}{\gamma_i}.\]
%
%
 Considering the constraints $p_i \geq 0$, the optimum powers are given explicitly by 
\[
	p_i^* =  \left\{
	\begin{array}{rl} 
		\frac{1}{\gamma_i} \MMSE_i^{-1} 
		\left(\zeta_i\right) \quad & \zeta_i < 1 \\
		0 \quad & \zeta_i \geq 1
	\end{array} \right.
	\]
 where $\zeta_i = \lambda/\gamma_i$. This solution has a graphical interpretation analogous to conventional water-filling \cite{Lozano2006} with $\frac{1}{\gamma_i}$  exchanged for $\frac{\Gamma_i(\zeta_i)}{\gamma_i},$ where
\[
	\Gamma_i(\zeta_i) = \left\{
	\begin{array}{rl} 
		1/\zeta_i - \MMSE_i^{-1}(\zeta_i) \quad & \zeta_i < 1 \\
		1 \quad & \zeta_i \geq 1
	\end{array} \right.
	\]
is the gap with respect to an ideal Gaussian signal. For Gaussian inputs, $\Gamma_i = 1$. 

The $\lambda$ that maximizes the EE is obtained by finding the root of $F(\lambda)$. The rate functions $r_i$ are computed through integration of the MMSE over $\rho$ \cite{Guo2005}, \[r_i(\rho_i) = \int_0^{\rho_i} \MMSE_i(\rho) d\rho.\] As already mentioned, the MMSE can be evaluated for discrete constellations in a semi-analytical form involving some simple integrals. For a real-time implementation the values of $\MMSE_i^{-1}(\cdot)$ and $r_i(\cdot)$ can be tabulated for the constellations of interest. The function $F(\lambda)$ is then evaluated as follows:
\begin{enumerate}
\item Calculate $\zeta_i$ for all subcarriers
\item Use the table of $\MMSE^{-1}(\cdot)$ to find $p_i^*$ 
for all subcarriers 
\item Use the table of $r_i(\cdot)$ to find $r_i(\gamma_i p_i^*)$ for all subcarriers
\item Use $r_i(\gamma_i p_i^*)$ and  $p_i^*$ to calculate $F(\lambda)$
\end{enumerate}

	\subsection{Nested convex problem}

 Many solutions (whether closed-form or algorithmic) to maximization of rate functions given a sum power constraint in various scenarios are available in the literature.
 A well-known example of this is rate maximization over parallel channels. 
 The solution is water-filling, where the water level is a function of the dual variable, which can be computed using known algorithms \cite{Palomar2005b}. An EE optimization problem can be reduced to a one-dimensional convex problem using transformation \eqref{eq:transformation}, which allows the known results to be utilized.
We will illustrate this using the example of mercury/water-filling.

 For any optimization problem, we can first optimize over some of the variables and then over the remaining ones \cite[Sec. 4.1.3, p. 133]{Boyd2004}. Thus, \eqref{eq:Parameter-freeConvex} can be reformulated as
\begin{equation}
  \underset{t > 0}{\text{maximize}} \quad t f_1(\boldsymbol{y}^*(t)/t), 
\label{sec:nestedproblem}
\end{equation}
where 
\begin{alignat}{1}
	\boldsymbol{y}^*(t) & =
	\argmax_{\boldsymbol{y}}
	\left\{
  f_1(\boldsymbol{y}/t) :
  	t f_2(\boldsymbol{y}/t) \leq 1, 
  	\boldsymbol{y}/t \in \mathcal{S}
  		\right\} = t\boldsymbol{x}^*(t)  \nonumber \\
	& =	t\, \argmax_{\boldsymbol{x}}
	\left\{
  f_1(\boldsymbol{x}) :
  	f_2(\boldsymbol{x}) \leq 1/t, 
  	\boldsymbol{x} \in \mathcal{S}
  		\right\}.  		
\label{eq:nestedsolution}
\end{alignat}
Since the original problem is convex, the new problem is convex as well. 
 
 As shown in \cite{Lozano2006},
 the optimal power allocation for the maximization of the sum rate (or mutual information) over parallel channels for an arbitrary modulation scheme, \emph{i.e.} \[\boldsymbol{p}^* =
	 \underset{\substack{\boldsymbol{p} \geq \boldsymbol{0} \\ \sum_{i=1}^K p_i \leq P}}{\argmax} \sum_{i=1}^K r_i(p_i),\]
where $r_i(p_i)$ is given by \eqref{eq:mut_inf}, is \[p_i^*=\frac{1}{\gamma_i} \MMSE_i^{-1}
				\left( \min \left\{ 1,\frac{\eta}{\gamma_i} \right\}	\right),
				\quad i=1,\dots,K,\]
where $\eta$ is the unique solution to the equation 
\begin{equation}
	\sum_{i=1, \gamma_i>\eta}^K \frac{1}{\gamma_i} \MMSE_i^{-1}
		\left( \frac{\eta}{\gamma_i} \right) = P.
		\label{eq:nestedConstraint}
\end{equation}
 Let us denote the maximum rate function by $\sum_{i=1}^K r_i(p_i^*(P))$, which is evaluated algorithmically for any given sum power $P\geq0$. Now we want to solve the problem
 \[
     \underset{\boldsymbol{p} \geq \boldsymbol{0}}{\text{maximize}}
     \frac {\sum_{i=1}^K r_i(p_i)}
                    {\mu + \sum_{i=1}^K p_i},\]
	with $\mu>0$.
  Applying \eqref{sec:nestedproblem} and the known solution $p_i^*$, we obtain
\begin{equation}
  \underset{0 < t \leq 1/\mu}{\text{maximize}} \quad t \cdot \sum_{i=1}^K r_i(p_i^*(1/t-\mu)), 
  \label{eq:nestedMWF}
\end{equation}
where $t = (\mu + P)^{-1}$.

 Note that the optimal power allocation for EE maximization is functionally identical to that of rate maximization. 
 The difference between them is that $\eta$ is chosen to fulfill the sum power constraint in the former, whereas $\eta$ is chosen to achieve the highest EE in the latter.

 A similar nesting approach was proposed in \cite{Chong2011a}, where the EE problem with any concave rate function is reduced to a one-dimensional quasiconvex problem. Here it is formulated as a one-dimensional convex problem. 
 
 This approach has the advantage that known rate maximization results can be easily implemented with almost no analysis required for maximizing the EE. However, doing some pre-analysis of the original EE optimization problem enables it to be solved with less computational cost.
 In solving \eqref{eq:nestedMWF}, every iteration for finding the optimal $t$ requires solving \eqref{eq:nestedConstraint} to obtain $p_i^*(1/t-\mu)$. 
 In the approach presented in Section \ref{sec:merc-waterf}, however, no inner optimization is required because $p_i^*$ is derived explicitly as a function of $\lambda$. 
 Thus, the optimization can be carried out directly over the dual variable $\lambda$ and the maximum EE is obtained more efficiently.
 On the other hand, if such a pre-analysis cannot be done,
 or if the computation time is not an essential criterion, the nesting method may be attractive.

\section{Simulation}
\label{sec:simulation}

\subsection{Time-varying channel with varying number of antennas}

\begin{figure}
	\centering
		\includegraphics[width=0.9\columnwidth]{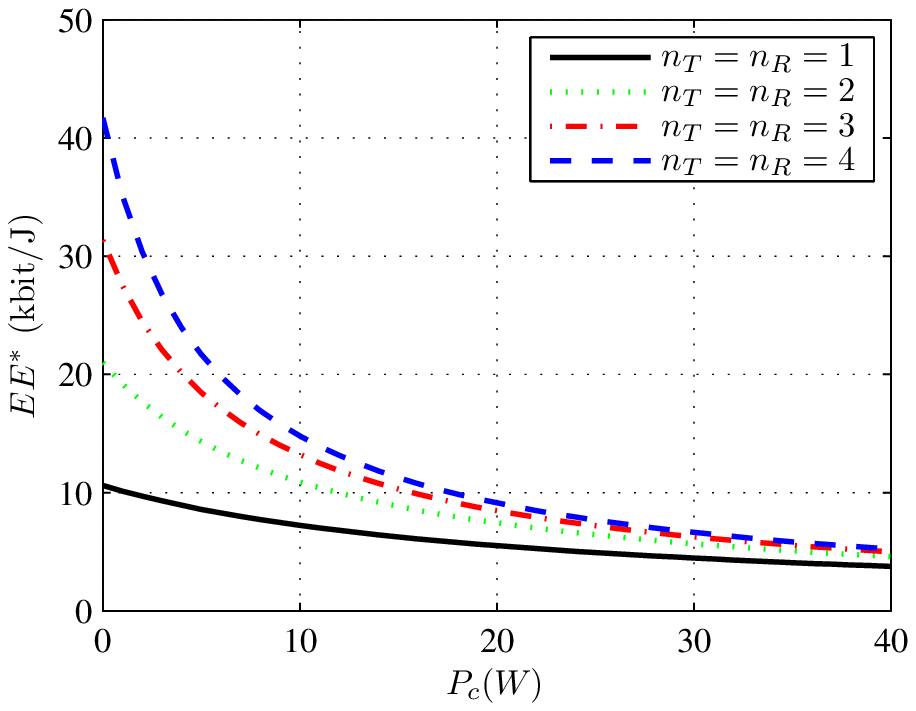}	
\caption{\label{fig:EE-ant-TR} The maximum EE in time-varying MIMO channels with Rayleigh fading versus circuit power. The number of transmit and receive antennas are identical.}
\end{figure}

\begin{figure}
\centering
\begin{overpic}[width=0.9\columnwidth,clip]{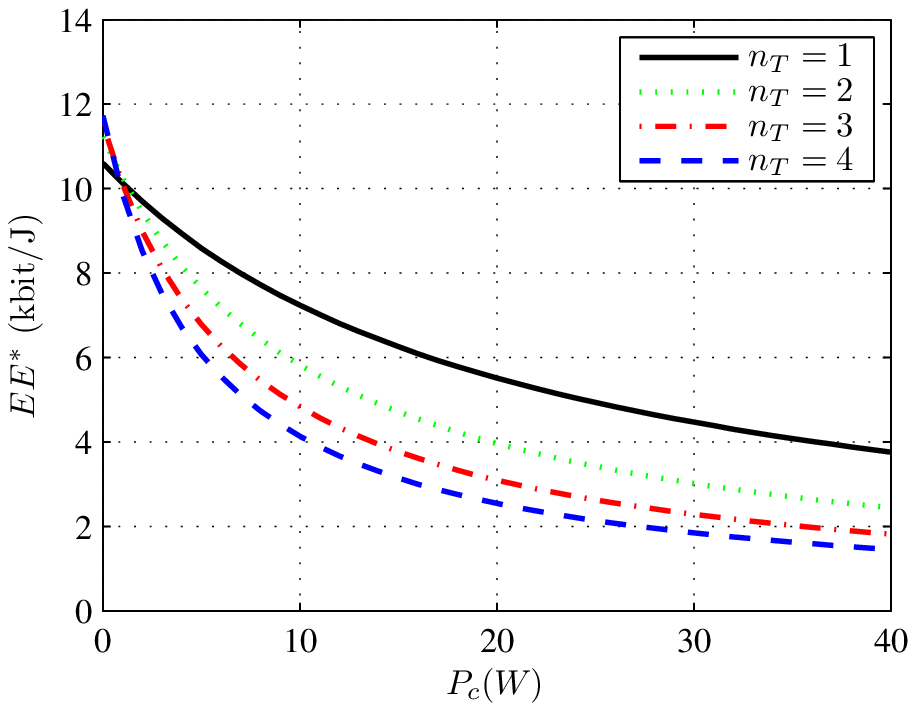}
		\put(120,570){
		\includegraphics[scale=0.33,clip]{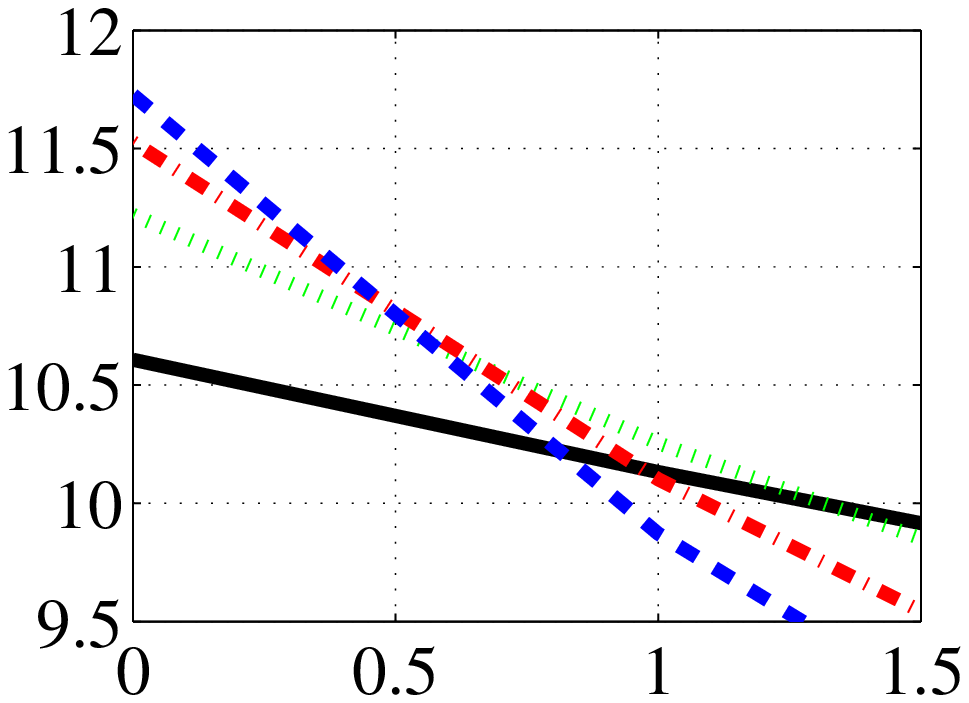}}	
\end{overpic}
\caption{\label{fig:EE-ant-TR1} The maximum EE in time-varying MIMO channels with Rayleigh fading versus circuit power. The number of receive antennas is one. The inset shows the enlarged region where $P_c\in [0,1.5]$.}
\end{figure}

 Let us consider a time-varying frequency-flat MIMO link with $n_T$ and $n_R$ transmit and receive antennas, respectively, where the link between each transmitter and receiver antenna is subject to Rayleigh fading.   
 We assume that perfect causal channel information is available at both ends. 
 As previously mentioned, this can be transformed to parallel channels using singular-value decomposition. 
 Using the result from Section \ref{sec:TVparchan} and the generic base station power model in Section \ref{sec:genericBS}, we optimize the EE over the transmit power for various antenna configurations and observe how the optimal EE changes with the circuit power $P_c$. 
 The bandwidth is set at $B=200$ kHz, and the noise power density at $N_0=-104.5$ dBm/Hz. 
 We assume the power amplifier efficiency to be $\eta_{\text{PA}}=0.35$.
 The other constants in the power model are chosen according to values presented in \cite{Chen2011}: 
$\eta_C=0.95$, 
$\eta_{\text{PS}}=0.9$, 
$P_{\text{sta}}=20$ W. 

 In Fig. \ref{fig:EE-ant-TR} we observe that for an equal number of antennas ($n_T=n_R=n$) on both ends, it is more efficient to employ more antennas in this setting.
 Notice also that $EE^*$ decreases monotonically with $P_c$. 
 This is in agreement with results in \cite{Prabhu2010a}, although there the antenna configuration is considered to be energy-efficient if it yields a small energy-per-goodbit given a maximum tolerated outage probability. 
 It is shown there that for Rayleigh fading, selecting the balanced MIMO configuration with the highest $n$ gives the best EE, but this is not the case for Rician fading. Due to higher correlation between the transmit and receive antennas in Rician fading, lower rates are achieved and therefore the employment of more antennas (which incur higher circuit power consumption) deteriorates the EE. 
  
 It is also interesting to note that if $P_c=0$, \emph{i.e.} if the circuit power does not depend on the number of antennas, $EE^*$ increases linearly with $n$.
 
 In Fig. \ref{fig:EE-ant-TR1}, we simulate the case where the receiver has only one antenna. 
 Again, $EE^*$ decreases with $P_c$. 
 However, it is not always best to choose the largest number of transmit antennas.
 As can be seen in the inset, employing the highest $n_T$ is efficient only if $P_c$ is small. 
 This is intuitive since when $P_c$ is small, it does not cost much more power to employ more antennas. 
 As $P_c$ increases, the loss in EE by employing more antennas increases as well.
 The reason for this is that when $n_R = 1$ and $P_c$ is nonzero, the transmission rate scales sublinearly with $n_T$, whereas the power consumption scales linearly with it. 
 As $P_c$ becomes larger, the difference between the gain in EE (through the increase of the transmission rate by increasing $n_T$) and the loss caused by the more rapid increase in power consumption becomes larger as well.
  
 The overall conclusion from the assessment in Figures \ref{fig:EE-ant-TR} and \ref{fig:EE-ant-TR1} is that one should carefully consider whether or not to activate each antenna with the required RF chain. As a rule of thumb it holds: activate additional antennas at the transmitter and receiver side only if it is worth it. Contrary to the traditional point of view, having more antennas is not always better. An additional diversity gain (Fig. \ref{fig:EE-ant-TR1}) does not always justify the additional energy consumption; it depends on the operating point. In contrast the additional degree of freedom or multiplexing gain in Fig. \ref{fig:EE-ant-TR} motivates the activation of more antennas.

\subsection{Quadratic $m$-QAM}

In the presence of Gaussian noise, the MMSE for an $m$-ary discrete constellation is 
\[\MMSE(\rho) = 1 - \frac{1}{\pi} \int \frac{\left|\sum_{l=1}^{m} q_l s_l e^{-|y - \sqrt{\rho} s_l|^2}\right|^2}{\sum_{l=1}^{m} q_l e^{-|y - \sqrt{\rho} s_l|^2}} dy,\]
where $q_l$ are probabilities and the integral is over the complex field.

For $m$-PAM, we have $q_l = 1/m$ and \[s_l \in \left\{(2l-1-m) \sqrt{\frac{3}{m^2 - 1}}\right\}.\] For even $m$, the corresponding $m$-QAM consists of two $m/2$-PAM constellations in quadrature, each with half the power. Writing $y$ as $y_I + jy_Q$, it can be shown that integration over the quadrature component $y_Q$ yields $\sqrt{\pi}$. Thus, for $m$-PAM we have
\[\MMSE(\rho) = 1 - \frac{1}{\sqrt{\pi}} \int_{-\infty}^\infty \frac{\left(\sum_{l=1}^{m} q_l s_l e^{-(y_I - \sqrt{\rho} s_l)^2}\right)^2}{\sum_{l=1}^{m} q_l  e^{-(y_I - \sqrt{\rho} s_l)^2}} dy_I.\]

\begin{figure}
    \centering
    \resizebox{0.9\columnwidth}{!}{\includegraphics{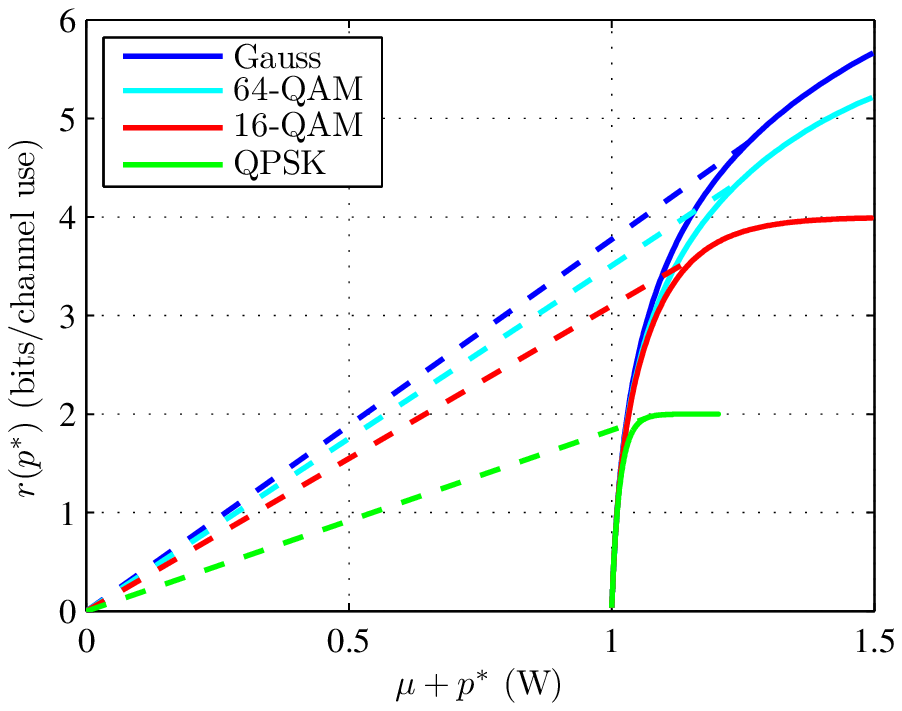}}
    \caption{Trade-off curve between transmit power and mutual information for Gaussian signals and $m$-QAM, respectively, in a single-carrier system with $\mu = 1$. The dotted lines indicate the maximum EE.}
    \label{fig:fig1_mqam}
\end{figure}

The values of $\MMSE(\rho)$ are evaluated numerically for various $m$-QAM constellations.
Using this result, $\MMSE^{-1}(\cdot)$ and $r_i(\cdot)$ are tabulated. The EE of a flat fading channel is optimized according to the method detailed in Section \ref{sec:merc-waterf}. The resulting trade-off curve with $\mu = 1$ is shown in Figure \ref{fig:fig1_mqam}. If $\mu$ is independent of the modulation scheme, it is always beneficial to use a higher modulation order since there is no cost associated with using a higher order modulation scheme. For small values of $\mu$, the curves start at a point close to the origin and the optimal EE is approximately equal for the different schemes, whereas the difference increases for larger values of $\mu$. The value of $p^*$ is also higher for higher-order modulation schemes.

However, a higher modulation scheme may increase the necessary offset power. In this case, a lower modulation order might be optimal in certain cases.

\section{Discussion}
\label{sec:discussion}

The variable $\lambda$ is found throughout the solutions in the application examples. We would like to point out its significance by recapitulating its various interpretations. In Section \ref{sec:par-convex-prog} we showed that $\lambda$ represents the relative weight of the denominator in the scalarized bi-criterion optimization problem. It can also be interpreted as the slope of the trade-off curve between two objectives. In EE optimization, these two objectives are the sum rate and the sum power. At the optimum, $\lambda^*$ is identical to the maximum EE adjusted with an appropriate system-dependent scaling factor.

All the examples we considered resulted in water-filling solutions. It is noteworthy that $\lambda$ in these cases represents a cut-off value, \emph{i.e.} power is allocated for transmission through a channel only if the SNR value $\gamma$ is larger than $\lambda$.

\section{Conclusions}
\label{sec:conclusions}

There exist many results on EE optimization in wireless communications systems. Most papers formulate a novel objective function and solve the corresponding optimization problem under certain constraints and assumptions for a specific scenario. We feel that it is time to unify the various approaches and understand the core of this class of problems. In this paper, motivated by a typical anecdotal scenario we arrive at a non-convex optimization problem of maximizing the ratio of achieved rate to dissipated power. It belongs to a class of problems called fractional programs, for which a rich but scattered mathematical literature has evolved over the years. Therefore, we collect and coherently present the results and offer a set of solution methods. The power models are carefully described in order to motivate the problem formulation. Applications in various settings include time-invariant parallel channels, time-varying flat-fading channels, and time-varying parallel channels, illustrating the usefulness of the framework. As an extension to this framework, one could study the case where more general function classes, e.g. non-concave functions, are used in the numerator of the EE metric. A framework that accommodates discrete optimization variables would also be interesting for systems with on-off power modes, in which parts of a base station may be turned off during off-peak hours. For these problems, other optimization methods will be needed in addition to concave fractional programming.
\section*{Acknowledgment}

The authors would like to thank their colleagues at Technische Universit\"at Dresden for various suggestions, especially Eckhard Ohlmer and Vinay Suryaprakash for critically reading the manuscript. This work was sponsored by the Federal Ministry of Education and Research (BMBF) within the scope of the Leading-Edge Cluster "Cool Silicon".

\bibliographystyle{IEEEtran}
\bibliography{referencesEE}

\begin{IEEEbiography}[{\includegraphics[width=1in,height=1.25in,clip,keepaspectratio]{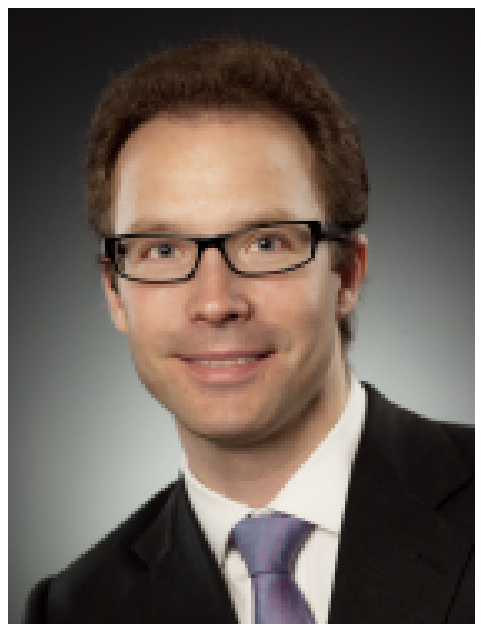}}]{Christian Isheden}
received his M.S. and Ph.D. degrees from Uppsala University and Royal Institute of Technology (KTH), both in Sweden, in 2000 and 2005, respectively. After working in various positions in the microelectronics industry, he joined the Vodafone Chair at Technische Universität Dresden, Germany, as a post-doctoral researcher in 2009. His research on energy-efficient link adaptation was recognized with the Best Paper Award at IEEE GLOBECOM 2010. In November 2011, he joined Actix GmbH in Dresden as a Senior Research Engineer. His current research interests include energy savings management and the coordination of SON use cases.

\end{IEEEbiography}
\begin{IEEEbiographynophoto}{Zhijiat Chong}
received his Dipl. Phys. degree in Physics from the Technical University of Dresden (TUD), Germany in 2009. In the same year, he joined the Chair of Communications Theory at TUD as a research associate, working on energy-efficient wireless communications in the project Cool Cellular within the frame of Cool Silicon. His current research interests include energy-efficient resource allocation and optimization.
\end{IEEEbiographynophoto}

\begin{IEEEbiography}[{\includegraphics[width=1in,height=1.25in,clip,keepaspectratio]{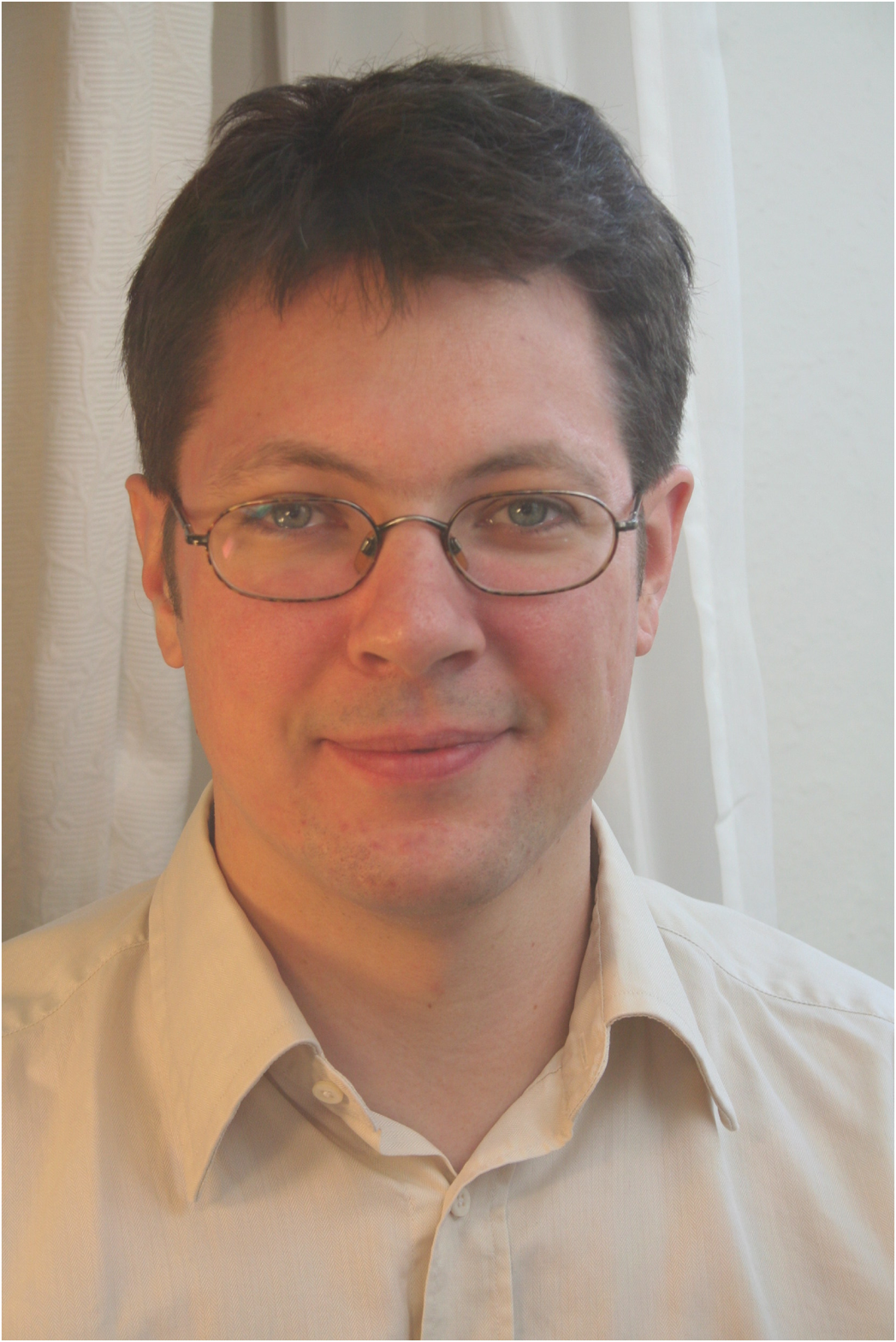}}]{Eduard A. Jorswieck}
received his Diplom-Ingenieur degree and Doktor-Ingenieur (Ph.D.) degree, both in electrical engineering and computer science from the Berlin University of Technology (TUB), Germany, in 2000 and 2004, respectively. He was with the Fraunhofer Institute for Telecommunications, Heinrich-Hertz-Institute (HHI) Berlin, from 2001 to 2006. In 2006, he joined the Signal Processing Department at the Royal Institute of Technology (KTH) as a post-doc and became a Assistant Professor in 2007. Since February 2008, he has been the head of the Chair of Communications Theory and Full Professor at Dresden University of Technology (TUD), Germany. His research interests are within the areas of applied information theory, signal processing and wireless communications. He is senior member of IEEE and elected member of the IEEE SPCOM Technical Committee. From 2008-2011 he served as an Associate Editor and since 2012 as a Senior Associate Editor for IEEE SIGNAL PROCESSING LETTERS. Since 2011 he serves as an Associate Editor for IEEE TRANSACTIONS ON SIGNAL PROCESSING. In 2006, he was co-recipient of the IEEE Signal Processing Society Best Paper Award. 
\end{IEEEbiography}
\begin{IEEEbiography}[{\includegraphics[width=1in,height=1.25in,clip,keepaspectratio]{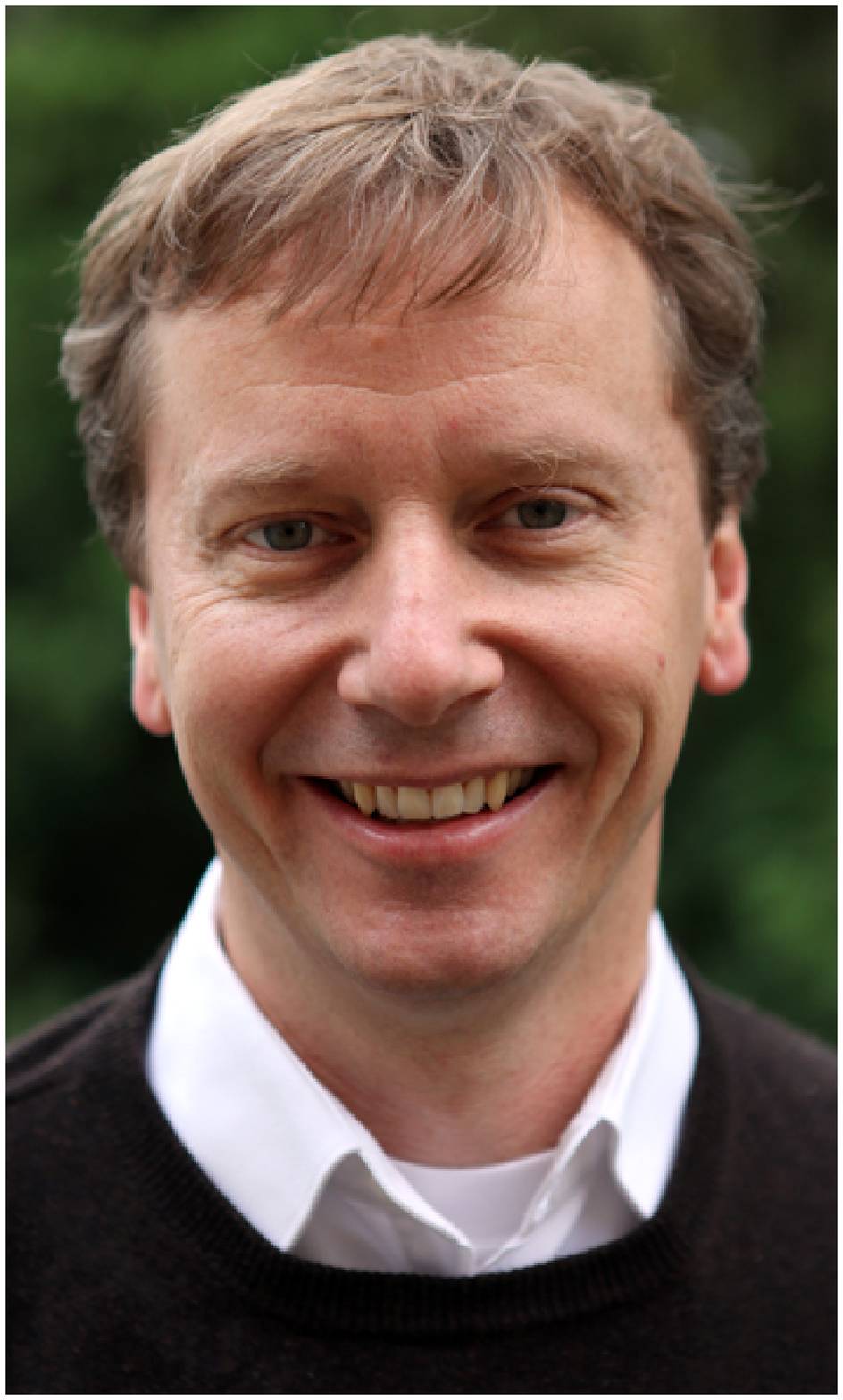}}]{Gerhard Fettweis}
earned his Ph.D. from RWTH Aachen (with H. Meyr) in 1990. Thereafter he was Visiting Scientist at IBM Research in San Jose, CA, working on disk drive read/write channels. From 1991-1994 he was Scientist with TCSI, Berkeley, CA, developing cellular phone chip-sets. Since 1994 he is Vodafone Chair Professor at TU Dresden, Germany, with currently 20 companies from Asia/Europe/US sponsoring his research. He runs the world's largest cellular research test-bed in downtown Dresden.
Gerhard is IEEE Fellow, Distinguished Speaker of IEEE SSCS, recipient of the Alcatel-Lucent Research Award and IEEE Millennium Medal. He has spun-out nine start-ups so far: Systemonic, Radioplan, Signalion, InCircuit, Dresden Silicon, Freedelity, RadioOpt, Blue Wonder Communications, INRADIOS.
Gerhard was TPC Chair of IEEE ICC 2009 (Dresden), and has organized many other events. He was elected Member-at-Large of IEEE SSCS (1999-2004) and COMSOC (1998-2000). He served as Associate Editor for IEEE JSAC (1998-2000) and IEEE Transactions CAS-II (1993-1996). 1991-1998 he was COMSOC's delegate within the IEEE Solid State Circuits Council. Gerhard is member of COMSOC's Awards Standing Committee and the IEEE Fellow Committee, and is active in COMSOC Technical Committees (Communication Theory, Wireless). During 2008-2009 he chaired the Germany Chapter of IEEE IT Society. 
\end{IEEEbiography}

\end{document}